\documentclass[12pt, a4paper]{article} %scrartcl

\usepackage{setspace}

\usepackage[font=footnotesize,labelfont=bf]{caption}

\usepackage[authoryear]{natbib}
\bibpunct{(}{)}{;}{a}{,}{,}
\usepackage[english]{babel}
\usepackage[latin1]{inputenc}
\usepackage[T1]{fontenc}
\usepackage{lmodern}
\usepackage{amssymb}
\usepackage{xcolor,xspace,caption}
\usepackage{lscape}
\usepackage{multirow}
\usepackage{enumerate}
\usepackage{amsmath}

\usepackage[demo]{graphicx}
\usepackage{blindtext}

\usepackage{tablefootnote}

\usepackage{listings}			% Used for Code environments

\usepackage{titlesec}
\usepackage[autostyle]{csquotes}

\usepackage{longtable}

\usepackage{subcaption}

\definecolor{hint}{RGB}{191,63,0}

\definecolor{brightpink}{rgb}{1.0, 0.33, 0.64}

\usepackage{hyperref}
\hypersetup{linkcolor={blue}}

\newcommand{\IF}{\mathbf{I}}

\author{Lining Yu\thanks{Research associate at Ladislaus von Bortkiewicz Chair of Statistics, C.A.S.E. - Center for Applied Statistics and Econometrics, IRTG 1792, Humboldt-Universit\"{a}t zu Berlin, Unter den Linden 6, 10099 Berlin, Germany. Email: wsqiuning@hotmail.com.}\and Wolfgang Karl H\"{a}rdle\thanks{Ladislaus von Bortkiewicz Professor of Statistics, C.A.S.E. - Center for Applied Statistics and Econometrics, Humboldt-Universit\"{a}t zu Berlin, Unter den Linden 6, 10099 Berlin, Germany; Wang Yanan Institute for Studies in Economics, Xiamen University, 422 Siming Road, Xiamen 361005, China; Sim Kee Boon Institute for Financial Economics, Singapore Management University, 90 Stamford Road, Singapore 178903, Singapore; Department of Mathematics and Physics, Charles University Prague, Ke Karlovu 2027/3, 12116 Praha 2, Czech. Email: haerdle@wiwi.hu-berlin.de.} \and Lukas Borke \thanks{Research associate at Ladislaus von Bortkiewicz Chair of Statistics, C.A.S.E. - Center for Applied Statistics and Econometrics, Humboldt-Universit\"{a}t zu Berlin, Unter den Linden 6, 10099 Berlin, Germany. Email: lukas@borke.net.} \and Thijs Benschop \thanks{Research associate at Ladislaus von Bortkiewicz Chair of Statistics, C.A.S.E. - Center for Applied Statistics and Econometrics, Humboldt-Universit\"{a}t zu Berlin, Unter den Linden 6, 10099 Berlin, Germany. Email: thijs.benschop@hu-berlin.de.}}

\date{\;}

\usepackage[top = 1in, bottom = 1in, left = 1in, right = 1in]{geometry}

\begin{document}

\title{An AI approach to measuring financial risk \thanks{Financial support from the Deutsche Forschungsgemeinschaft
(DFG) via SFB 649 \enquote{\"{O}konomisches Risiko}, IRTG 1792 \enquote{High-Dimensional Non-Stationary Times Series} and Sim Kee Boon Institute for Financial Economics, Singapore Management University,as well as the Czech Science Foundation under grant no. 19-28231X, the Yushan Scholar Program and the European Union's Horizon 2020 research and innovation program ''FIN-TECH: A Financial supervision and Technology compliance training programme'' under the grant agreement No 825215 (Topic: ICT-35-2018, Type of action: CSA), Humboldt-Universit\"at zu Berlin, is gratefully acknowledged.
}}\maketitle

\begin{abstract}
\noindent AI artificial intelligence brings about new quantitative techniques to assess the state of an economy. Here we describe a new measure for systemic risk: the Financial Risk Meter (FRM).
This measure is based on the penalization parameter ($\lambda$) of a linear quantile lasso regression.
The FRM is calculated by taking the average of the penalization parameters over the 100 largest US publicly traded financial institutions.
We demonstrate the suitability of this AI based risk measure by comparing the proposed FRM to other measures for systemic risk, such as VIX, SRISK and Google Trends.
We find that mutual Granger causality exists between the FRM and these measures, which indicates the validity of the FRM as a systemic risk measure.
The implementation of this project is carried out using parallel computing, the codes are published on www.quantlet.de with keyword
\raisebox{-1.5pt}{\includegraphics[scale=0.2]{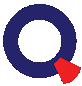}}\,{\href{https://github.com/Quantlet/FRM}{FRM}}.
The \textsf{R} package \textbf{RiskAnalytics} is another tool with the purpose of integrating and facilitating the research, calculation and analysis
methods around the FRM project.
The visualization and the up-to-date FRM can be found on \href{http://frm.wiwi.hu-berlin.de/}{hu.berlin/frm}.
%\url{hu.berlin/frm}
\end{abstract}



\textit{This is a post-peer-review, pre-copyedit version of an article published in the Singapore Economic Review. The final authenticated version is available online at:}\\ \url{http://dx.doi.org/10.1142/S0217590819500668}

%%%%%%%%%%%%%%%%%%%%%%%%%%%%%%%%%%%%%%%%%%%%%%%%%%%%%%%%%%%%%%%%%%%%%%%%%%%%%%%%%%%%%%%%%%
%%%%%%%%%%%%%%%%%%%%%%%%%%%%%%%%%%%%%%%%%%%%%%%%%%%%%%%%%%%%%%%%%%%%%%%%%%%%%%%%%%%%%%%%%%
\section{Introduction}

Systemic risk is hazardous for the stability of financial markets, since the failure of one firm may impact the stability of the whole system.
There are various definitions of systemic risk. One of the most popular definitions is introduced in \cite{Steven2008}.
He defined systemic risk as a trigger event, such as an economic shock or institutional failure, causing a chain of bad economic consequences, sometimes referred
to as domino effect. This definition indicates that interlinkages and interdependencies in a system or market are very crucial for controlling systemic risk.
The financial crisis in 2008 is an example. After the bankruptcy of Lehman Brothers, several more financial cooperations bankrupted as a result of their
interlinkages with Lehman Brothers. Consequently, there has been a surge in the interest in measuring and controlling systemic risk since the 2008 crisis,
which has led to an increase in the research on this topic.
\vspace{0.1cm}

Several methodologies for measuring systemic risk have been proposed. \cite{Adrian:2016} proposed CoVaR, the value at risk of financial institutions conditional on
the other institutions being under distress, which uses two linear quantile regressions.
\cite{Hautsch:2015} refined this algorithm by introducing linear quantile lasso regression with a fixed penalization parameter $\lambda$ for each company to select
the relevant risk drivers. \cite{Fan:2015} and \cite{HWY2016} use a nonlinear Single Index Model (SIM) combined with a variable selection technique to select
risk factors. In their application, they use data on 200 financial companies and 7 macro variables to estimate CoVaR. During the estimation procedure, a time-varying penalization parameter $\lambda$ is generated. This series has a striking pattern: higher values correspond to financial crisis times and lower values
correspond to stable periods. This observation has led to the idea to use the penalization parameter $\lambda$ itself as a measure for systemic risk.
The time-varying feature is not observed in \cite{Hautsch:2015}, who applied a fixed $\lambda$ for each firm, and not time varying.
\vspace{0.1cm}

\cite{Fan:2015} provide the $\lambda$ series for single companies. In contrast, we would like to see the overall behavior of $\lambda$.
\cite{HWY2016} compare the linear quantile lasso model and SIM, and conclude that SIM is more suited than the linear model, but that the linear
quantile lasso model is also valid in terms of backtesting. Indeed if one generates $\lambda$ series for 100 firms with more than $300$ observations each, then the application of SIM is not realistic. Since linear quantile lasso is easier to apply and time saving, we decided to use it to compute the FRM.
\vspace{0.1cm}

We use log return data from the 100 largest US publicly traded financial institutions as well as 6 macro variables. Our model is based on daily log returns of these
financial institutions. The time period under consideration runs from April 5, 2007 until September 23, 2016 and covers several documented financial crises
(2008, 2011). We observe that the pattern of this risk measure is more informative on financial risk.
The shape and volatility of the series correspond to the market volatility and financial events with a large impact on systemic risk are clearly visible. Therefore,
this series of averaged $\lambda$ may be called a Financial Risk Meter (FRM). \cite{SFB649DP2016-047} apply linear quantile lasso regression to analyze the behavior of the $\lambda$ series. They find that $\lambda$ is
sensitive to the changes of volatility, which provide the theoretical evidence for the FRM to be a systemic risk measure, as high volatility indicates high risk.
\vspace{0.1cm}

We introduce the methodology of the FRM, describe the risk levels, the computational implementation as well as possible visualizations.
We compare the FRM with other systemic risk measures, such as VIX \citep[][]{Norman2009}, SRISK \citep[see][]{Brownlees:2016} as
well as the Google trends of key words related to financial crises \citep[see][]{GT2013}. We find that the FRM and these risk measures mutually Granger cause,
which indicates the validity of the FRM as a systemic risk measure.
\vspace{0.1cm}

The remainder of this paper is organized as follows. In Section 2 the methodology used to construct our FRM, which is quantile lasso modeling, is presented.
Section 3 presents the data, computational challenge and the visualization of the results. Section 4 shows the validity of our FRM as a measure for financial risk
by comparing with other financial risk measures.
%Section \ref{RiskAnalytics} describes the \textsf{R} package \textbf{RiskAnalytics} \citep{RiskAnalytics} facilitating real-time processing of Nasdaq and Yahoo
%finance data and parallelized quantile lasso regression methods.
Section 5 concludes, the financial institutions applied in this paper is listed in Section 6 Appendix.
All the \textsf{R} programs for this paper can be found on www.quantlet.de \citep{QuantNetMiningDP2}.
\vspace{0.1cm}

%%%%%%%%%%%%%%%%%%%%%%%%%%%%%%%%%%%%%%%%%%%%%%%%%%%%%%%%%%%%%%%%%%%%%%%%%%%%%%%%%%%%%%%%%%
%%%%%%%%%%%%%%%%%%%%%%%%%%%%%%%%%%%%%%%%%%%%%%%%%%%%%%%%%%%%%%%%%%%%%%%%%%%%%%%%%%%%%%%%%%
\section{FRM methodology and estimation}
In this section we describe the methodology and algorithm used to compute FRM. Since the penalization parameters are computed based on an $L_1$-norm (LASSO) quantile linear regression, this
regression framework is introduced first. Within this framework, the penalization parameter $\lambda$ is exogenous. Since the FRM is distilled from the selected penalization parameter, we subsequently discuss methods to select $\lambda$.
\vspace{0.1cm}

%%%%%%%%%%%%%%%%%%%%%%%%%%%%%%%%%%%%%%%%%%%%%%%%%%%%%%%%%%%%%%%%%%%%%%%%%%%%%%%%%%%%%%%%%%
\subsection{Linear Quantile Lasso Regression Model}
Following \cite{HWY2016}, we introduce the quantile lasso regression model. Let $m$ be the number of macro variables describing the state of the economy,
$k$ the number of firms under consideration, $j\in \{1,\ldots, k\}$. Then $p = k + m -1$ represents the number of covariates. $t \in \{1,\ldots, T\}$ is the
%time point with $T$ the total number of observations (days). $s$ is the index of moving window, $s \in \{1,\ldots,(T-n)\}$, where $n$ is the length of window size.
time point with $T$ the total number of observations (days). $s$ is the index of moving window, $s \in \{1,\ldots,(T-(n-1))\}$, where $n$ is the length of window size.
Then the quantile lasso regression is defined as:
\begin{equation}
X^s_{j,t}=\alpha^s_j+A_{j,t}^{s,\top}\beta^s_{j}+\varepsilon^s_{j,t},
\end{equation}
where $A^s_{j,t} \stackrel{def}{=}
$
$\left[
\begin{array}{c}
M^s_{t-1}\\
X^s_{-j, t}\\
\end{array}
\right]$, $M^s_{t-1}$ the $m$ dimensional vector of macro variables, $X^s_{-j, t}$ is the $p-m$ dimensional vector of log returns of all other firms except firm $j$
at time $t$ and in moving window $s$, $\alpha^s_j$ is a constant term and $\beta^s_{j}$ is a $p\times 1$ vector defined for moving window $s$.\\

The regression is performed using $L_1$-norm quantile regression proposed by \cite{Li:2008}:
\begin{eqnarray}
\label{loss:lambda}
\hspace{-0.4cm} \min\limits_{\alpha^s_j, \beta^s_j}\left\{n^{-1}
%\sum\limits_{t=s}^{s+n}\rho_{\tau}\big(X^s_{j,t}-\alpha^s_j-
\sum\limits_{t=s}^{s+(n-1)}\rho_{\tau}\big(X^s_{j,t}-\alpha^s_j-
A_{j,t}^{{s,\top}}\beta^s_{j}\big)
+ \lambda^s_j \parallel \beta^s_{j} \parallel_1\right\},
\end{eqnarray}

where $\lambda^s_j$ is the penalization parameter, and the check function $\rho_{\tau} (u)$ is:
\begin{equation*}
\rho_{\tau} (u) = |u|^{c} |\IF (u \leq 0) -\tau|,
\end{equation*}
where $c = 1$ corresponds to quantile regression. The $L_1$-norm quantile linear regression can be used to select relevant covariates (other firms and macro state
variables) for each firm.

%%%%%%%%%%%%%%%%%%%%%%%%%%%%%%%%%%%%%%%%%%%%%%%%%%%%%%%%%%%%%%%%%%%%%%%%%%%%%%%%%%%%%%%%%%
\subsection{Penalization Parameter $\lambda$}
\label{gacv}
Since Equation (\ref{loss:lambda}) has an $L_1$ loss function and an $L_1$-norm penalty term, the optimization problem is an $L_1$-norm quantile regression
estimation problem. The choice of the penalization parameter $\lambda_j^s$ is crucial. There are several options to select $\lambda_j^s$, e.g. with the
Bayesian Information Criterion (BIC) or using  the Generalized Approximate Cross-Validation criterion (GACV). \cite{Yuan:2006} conducted simulations and concluded
that GACV outperforms BIC in terms of statistical efficiency. Therefore, we determine $\lambda_j^s$ with the GACV criterion in the FRM model and set $\lambda_j^s$
as the solution of the following minimization problem:
\begin{equation*}
\min GACV(\lambda_j^s)=\min \frac{\sum_{t=s}^{s+(n-1)}\rho_{\tau}\big(X^s_{j,t}-\alpha^s_j-
A_{j,t}^{s \top}\beta^s_{j}\big)}{n-df},
\end{equation*}
where $df$ is a measure of the effective dimensionality of the fitted model.
The advantage of GACV is that it also works for $p>n$, which can be important for the FRM if the moving window size is small.
\vspace{0.1cm}

To compute the FRM, we perform the regression analysis as described above and select the $\lambda_{j}^{s,*}$ for each firm $j$
using GACV. The Financial Risk Meter
is defined as the average lambdas over the set of $k$ firms for all windows:

\begin{equation*}
FRM \stackrel{def}{=} \frac{1}{k} \sum_{j=1}^{k} \lambda_{j}^*
\end{equation*}

%%%%%%%%%%%%%%%%%%%%%%%%%%%%%%%%%%%%%%%%%%%%%%%%%%%%%%%%%%%%%%%%%%%%%%%%%%%%%%%%%%%%%%%%%%
%%%%%%%%%%%%%%%%%%%%%%%%%%%%%%%%%%%%%%%%%%%%%%%%%%%%%%%%%%%%%%%%%%%%%%%%%%%%%%%%%%%%%%%%%%
\section{Computational challenges and visualization}
%%%%%%%%%%%%%%%%%%%%%%%%%%%%%%%%%%%%%%%%%%%%%%%%%%%%%%%%%%%%%%%%%%%%%%%%%%%%%%%%%%%%%%%%%%
\subsection{Data}
\label{sec:data}
To compute the FRM, we use data from 100 US publicly traded financial institutions as well as six macro variables. The selection of financial companies is based
on the NASDAQ company list\footnote{See the NASDAQ webpage:\\http://www.nasdaq.com/screening/companies-by-industry.aspx?industry=Finance} and based on the market
capitalization. The selected companies are the 100 US publicly traded financial institutions with the largest market capitalization, see Table
\ref{table:List_of_Firms} in Appendix.
\vspace{0.1cm}

\begin{figure}[ht]
	\begin{center}
		\includegraphics[width=1.0\textwidth]{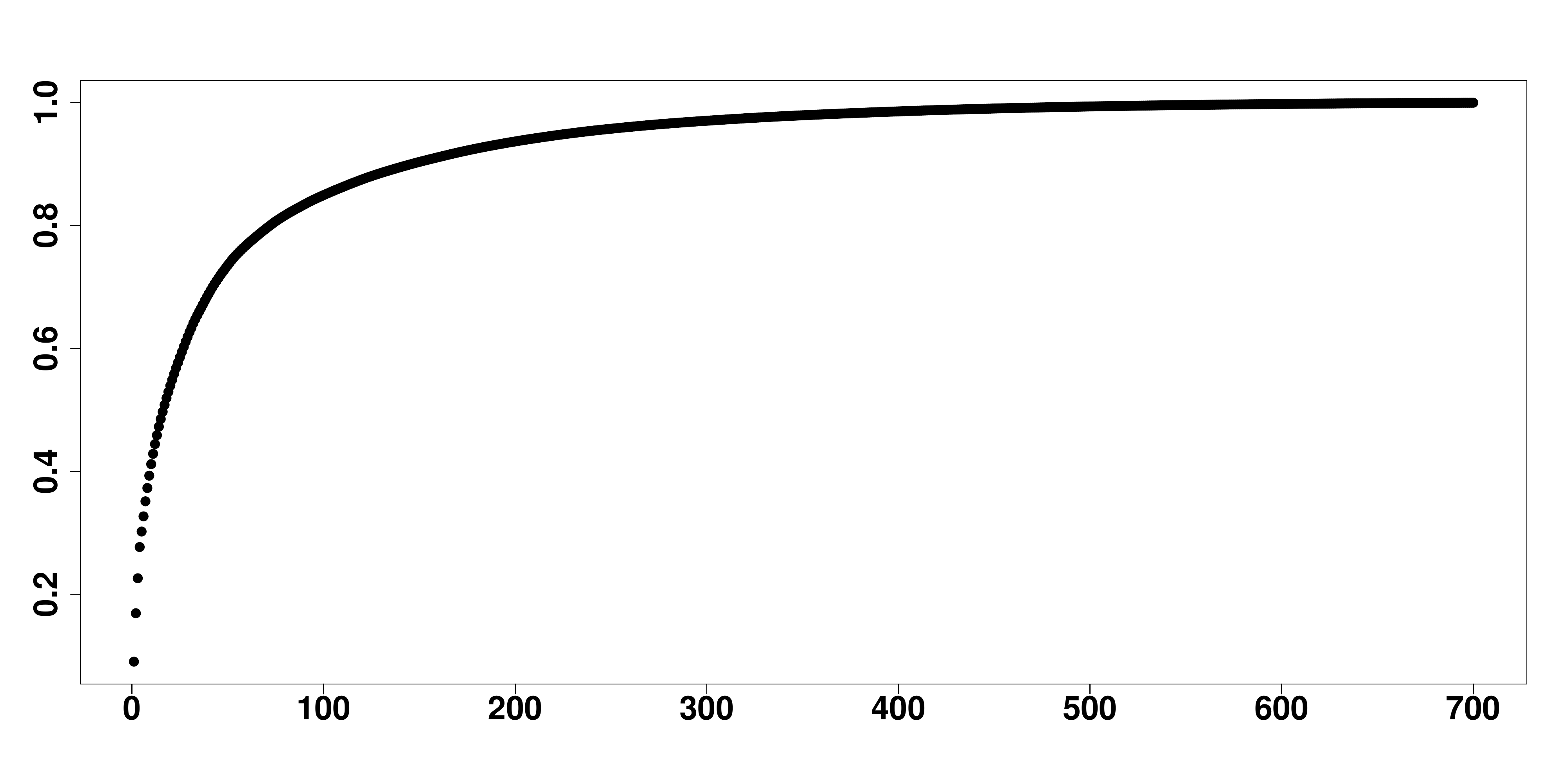}
  \end{center}
  \caption{The $x$-axis represents the number of firms ordered by market capitalization and the $y$-axis the percentage of total market capitalization.}
	\label{f7}
	\vspace{-0.5cm}
	\begin{flushright}
		\raisebox{-1.5pt}{\includegraphics[scale=0.2]{Figures/QLogo}}\,{\href{https://github.com/QuantLet/FRM/tree/master/FRM_per_cap}{FRM\_per\_cap}}
	\end{flushright}
\end{figure}

\begin{figure}[ht]
	\begin{center}
		\includegraphics[width=1.0\textwidth]{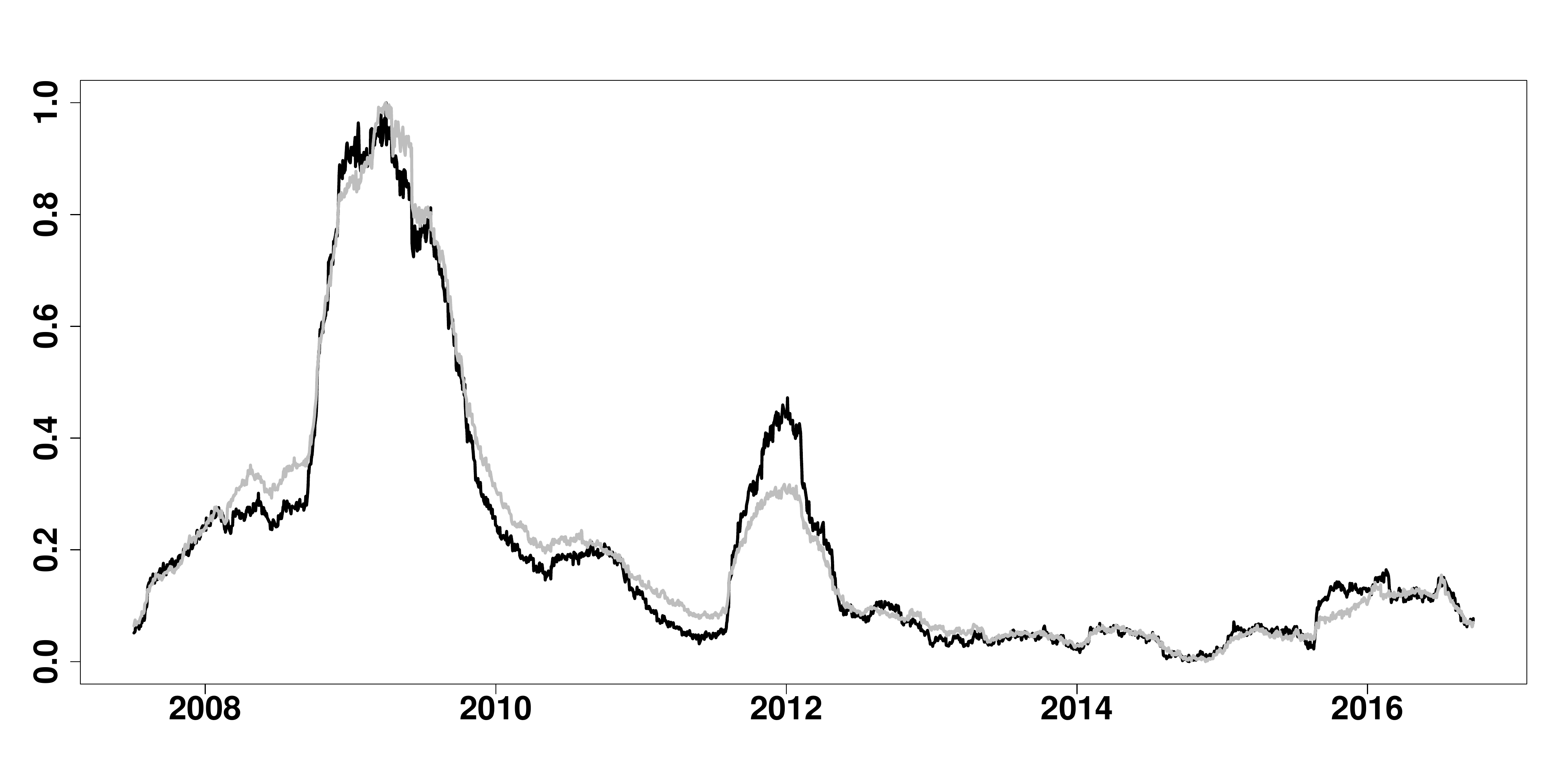}
	\end{center}
	\caption{FRM with 100 firms (black) and FRM with 200 firms (grey), moving window size $n = 126$.}
	\label{f11}
	\vspace{-0.5cm}
	\begin{flushright}
		\raisebox{-1.5pt}{\includegraphics[scale=0.2]{Figures/QLogo}}\,{\href{https://github.com/QuantLet/FRM/tree/master/FRM_compare_nf}{FRM\_compare\_nf}}
	\end{flushright}
\end{figure}

Initially, we used data on the 200 US publicly traded financial institutions with the largest market capitalization to compute the FRM. However, the smaller
companies in this set change regularly over the time period under consideration (2007-2016) due to, for instance, bankruptcies. This leads to obvious data download issues and therefore we use only 100 firms. Figure \ref{f7} shows the cumulative market capitalization of US financial firms. The $x$-axis
represents the firms ordered by market capitalization and the $y$-axis the cumulative market capitalization. We observe that the largest 100 firms cover more
than $85\%$ of the total market capitalization of all companies in the US financial market and we therefore can restrict our analysis to 100 firms.
Furthermore, the results of estimating the FRM based on 100 or 200 firms are very similar if the moving window size is the same. Figure \ref{f11} plots both
FRM series with the window size $n = 126$, the shape and the trends of them are similar.

%\vspace{0.5cm}

\begin{figure}[!ht]
	\centering
		\includegraphics[width=1.0\textwidth]{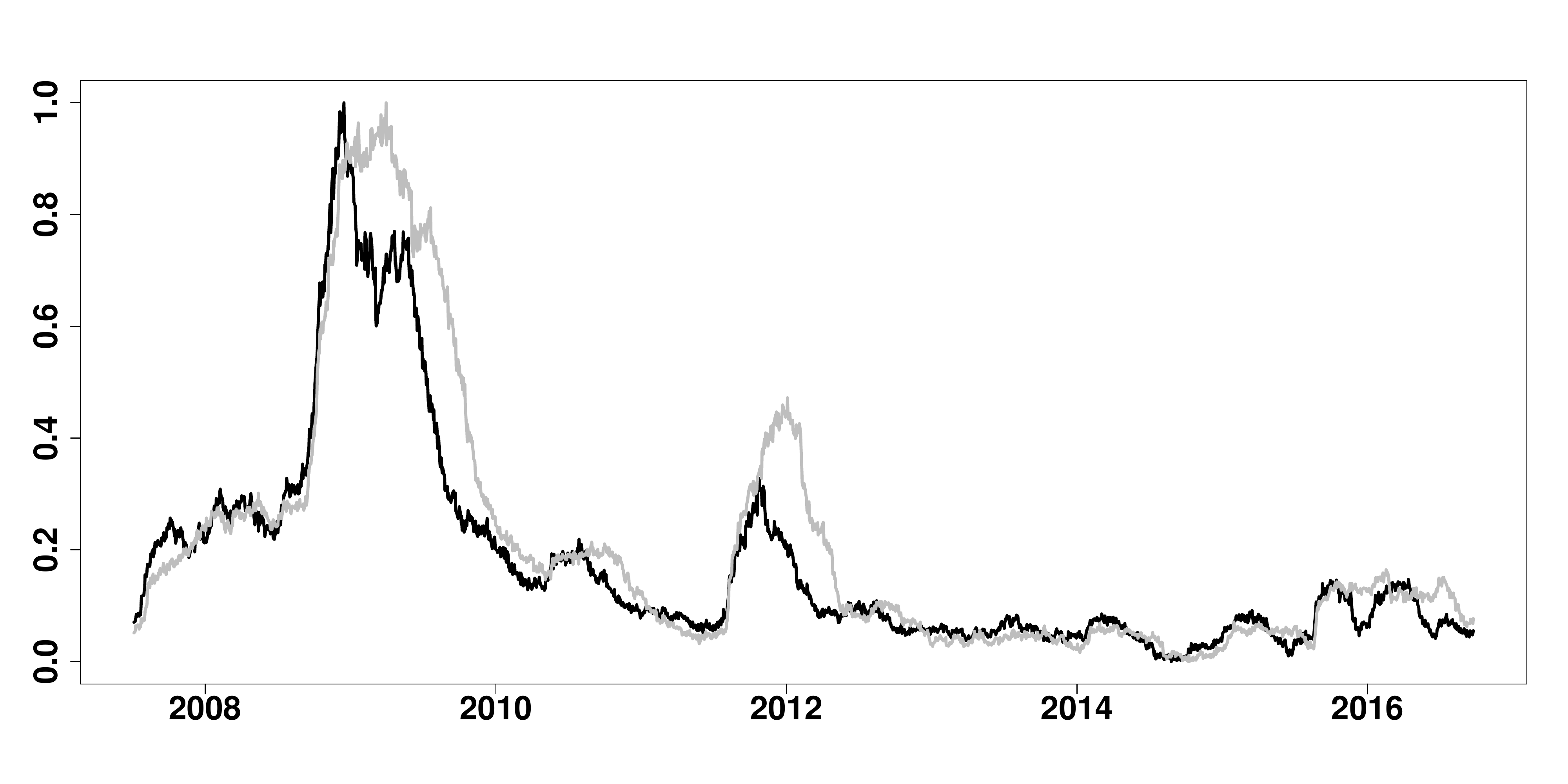}
		\caption{FRM with different moving window size, $n=63$ (black) and $n=126$ (grey), both series are scaled into the interval [0,1], from July 6, 2007
						 until September 23, 2016.}
		\label{f6}
		\vspace{-0.5cm}
		\begin{flushright}
			\raisebox{-1.5pt}{\includegraphics[scale=0.2]{Figures/QLogo}}\,{\href{https://github.com/QuantLet/FRM/tree/master/FRM_compare_ws}{FRM\_compare\_ws}}
		\end{flushright}
\end{figure}

We select six macro state variables to represent the general state of the economy: 1) the implied volatility index, VIX from Yahoo Finance;
(2) the changes in the three-month Treasury bill rate from the Federal Reserve Bank of St. Louis;
(3) the changes in the slope of the yield curve corresponding to the yield spread between the ten-year Treasury rate and the three-month bill rate from the
Federal Reserve Bank of St. Louis;
(4) the changes in the credit spread between BAA-rated bonds and the Treasury rate from the Federal Reserve Bank of St. Louis;
(5) the daily S\&P500 index returns from Yahoo Finance, and (6) the daily Dow Jones US Real Estate index returns from Yahoo Finance.
\vspace{0.1cm}

\begin{figure}[!ht]
	\centering
		\includegraphics[width=1.0\textwidth]{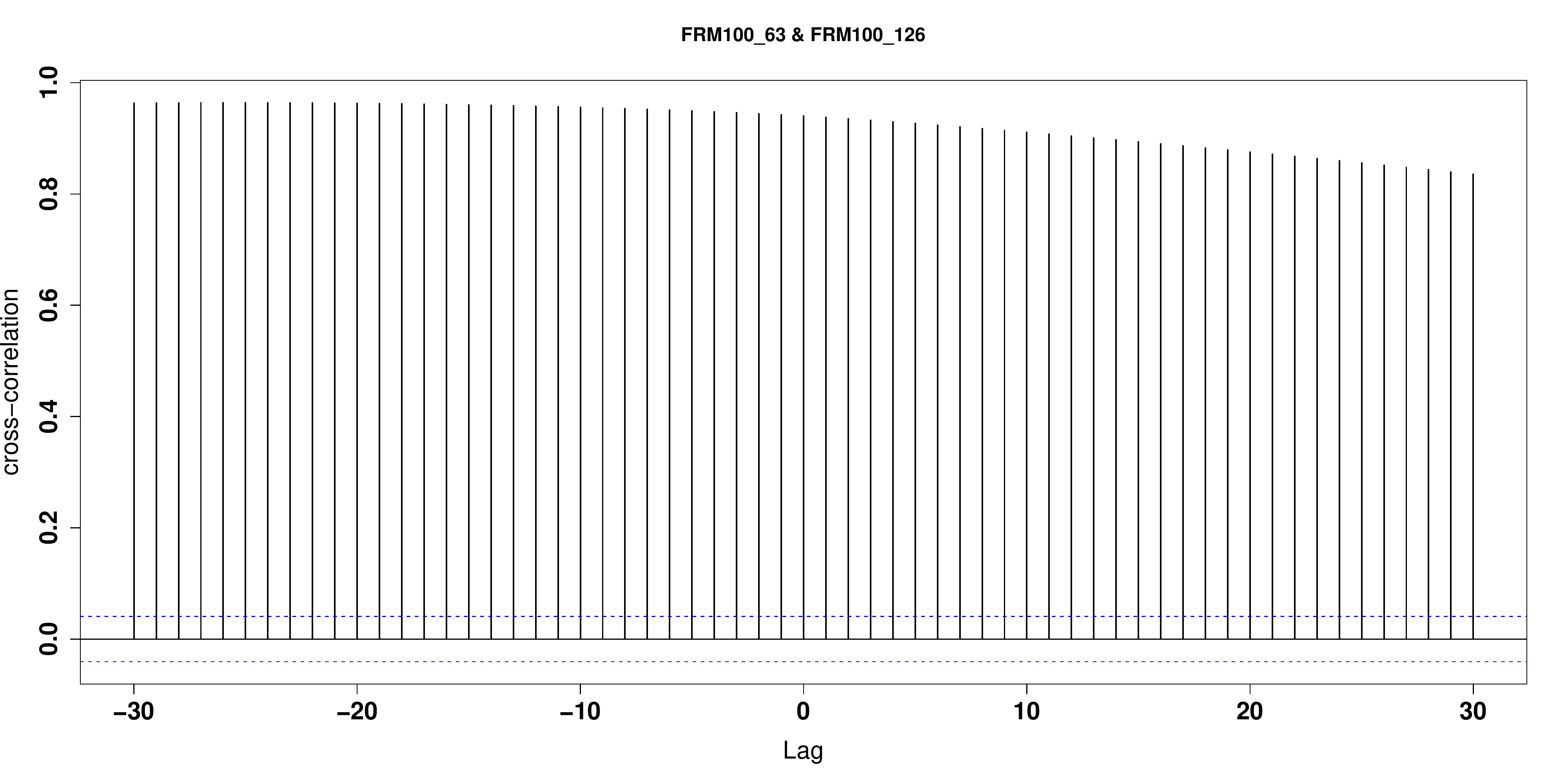}
		\caption{Cross correlation between FRM with $n = 63$ and FRM with $n = 126$, where the number of firms is 100.}
		\label{f13}
		\vspace{-0.5cm}
		\begin{flushright}
			\raisebox{-1.5pt}{\includegraphics[scale=0.2]{Figures/QLogo}}\,{\href{https://github.com/QuantLet/FRM/tree/master/FRM_compare_ws}{FRM\_compare\_ws}}
		\end{flushright}
\end{figure}

To compute the FRM we employ the tail parameter $\tau=0.05$. To find the a stable window size, $n$,
we have to make a trade-off. We find that the lasso selection technique performs worse if the window size is too small. Since we use daily data, the moving
window size should be larger than $50$, so that the estimation for each window is more precise.  The results of using different window sizes
(we have considered window sizes $n=63$ (one quarter) and $n=126$ (half a year)) are shown in Figure \ref{f6}. The larger the window size, the more lagged,
but also the smoother the plot is. Cross correlation can be used to determine the time delay of a time series, which we apply here for the estimate of the FRM
with $n=63$ and the FRM with $n=126$.  In Figure \ref{f13} and Table \ref{tb_12}, the largest autocorrelation between FRM with $n=63$ and the lagged FRM with
$n=126$ is $0.967$ from lag $-29$ to lag $-22$. We conclude that the FRM with $n=63$ leads the FRM $n=126$ by at least 22 periods. From all the preceding we set
the moving window size to $n = 63$.

\vspace{-0.5cm}

\begin{figure}[!ht]
	\centering
		\includegraphics[width=1.0\textwidth]{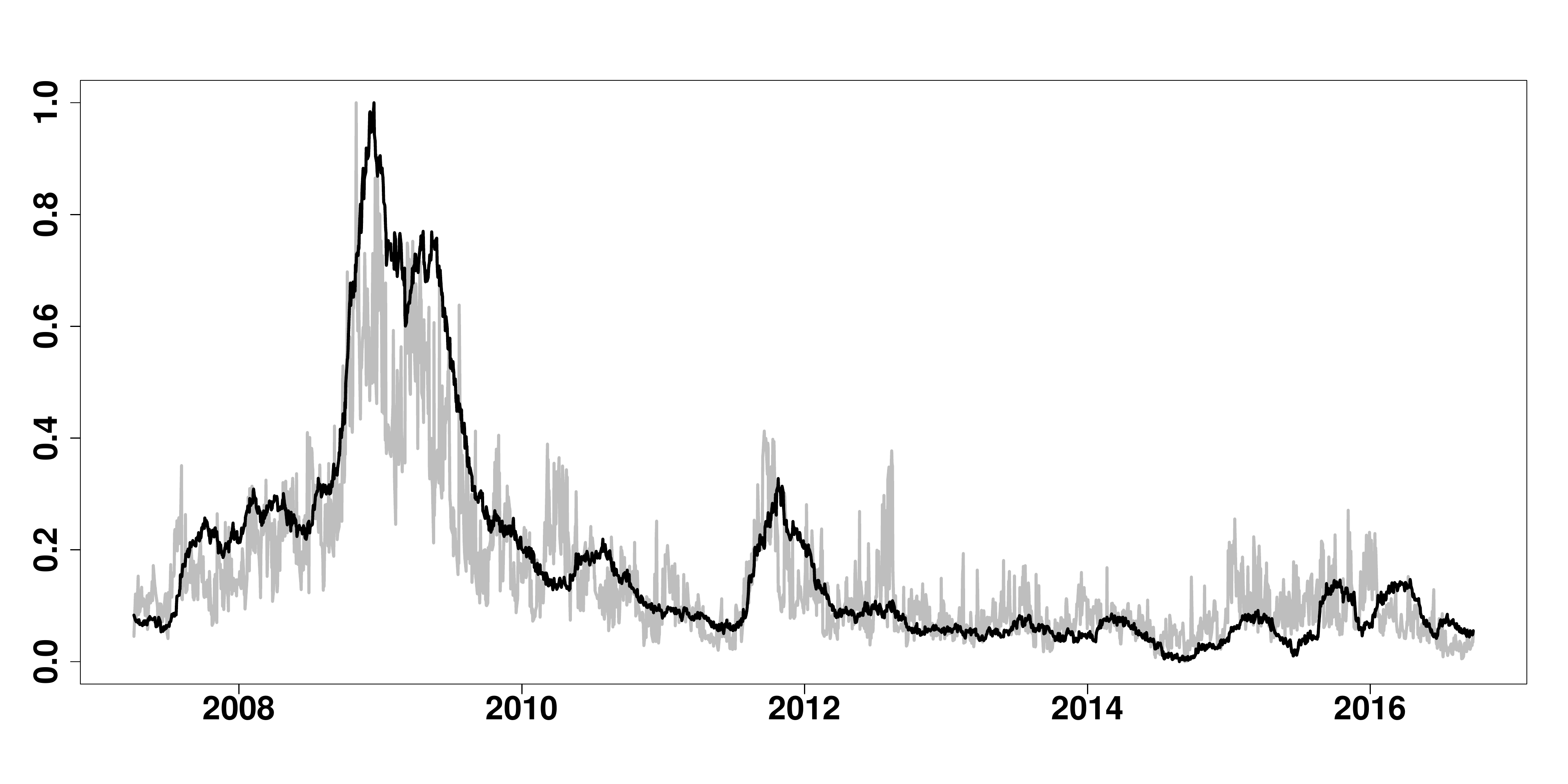}
		\caption{FRM (black) and $\lambda$ of Wells Fargo (grey), both series are scaled into interval [0,1], from April 5, 2007 until September 23, 2016.}
		\label{f10}
		\vspace{-0.5cm}
		\begin{flushright}
			\raisebox{-1.5pt}{\includegraphics[scale=0.2]{Figures/QLogo}}\,{{\href{https://github.com/QuantLet/FRM/tree/master/FRM_compare_of}{FRM\_compare\_of}}}
		\end{flushright}
\end{figure}

For each firm we have $2,386$ daily observations and $105$ covariates (99 firms and 6 macro-state variables). The FRM is the average of the $\lambda$'s computed
from the 100 individual firms. The $\lambda$'s for the individual firms are more volatile and less smooth than the average over 100 firms and therefore more robust
to reflect the impact from financial events on systemic risk. Figure \ref{f10} illustrates this by plotting the $\lambda$ of firm Wells Fargo (the largest firm
by market capitalization) and the FRM.

\renewcommand\arraystretch{1.5}
\begin{table}[!ht]
	\resizebox{\columnwidth}{!}{
	\begin{tabular}{ccccccccccc}
		\hline
		\hline
			Lag & -30& -29&  -28&  -27&  -26&  -25& -24& -23& -22& -21\\
		\hline
		Cross correlation & 0.963& 0.964& 0.964& 0.964& 0.964& 0.964 & 0.964& 0.964&  0.964& 0.963\\
		\hline
		\hline
	\end{tabular}}
  \caption{Cross correlation between the estimates of the FRM with $n = 63$ and FRM with $n = 126$.}
	\label{tb_12}
\end{table}

%%%%%%%%%%%%%%%%%%%%%%%%%%%%%%%%%%%%%%%%%%%%%%%%%%%%%%%%%%%%%%%%%%%%%%%%%%%%%%%%%%%%%%%%%%
\subsection{Computational challenges}
We wrote a script to automatically download the data from Yahoo Finance and Federal Reserve Bank of St. Louis. The {\textsf{R} package \emph{quantmod}
is used. More details and the script are available from Quantnet
(\raisebox{-1.5pt}{\includegraphics[scale=0.2]{Figures/QLogo}}\,{\href{https://github.com/QuantLet/FRM/tree/master/FRM_download_data}{FRM\_download\_data}}).
\vspace{0.1cm}

The $L_1$-norm quantile regression used to generate the $\lambda$ series is computationally intensive and therefore time-consuming, if applied sequentially for
a large number of firms, see for instance the code from Quantnet
(\raisebox{-1.5pt}{\includegraphics[scale=0.2]{Figures/QLogo}}\,{\href{https://github.com/QuantLet/FRM/tree/master/FRM_lambda_series}{FRM\_lambda\_series}}).
Therefore, we consider parallel computing in {\textsf{R} to reduce the computation time. {\textsf{R} offers several algorithms for performance
computing, such as \emph{lapply}, \emph{mclapply}, \emph{parLapply}, \emph{for} and \emph{foreach}
\footnote{The webpage http://www.parallelr.com/r-with-parallel-computing/ provides an overview.}. For our purposes the \emph{foreach} loops is the fastest solution,
which we use for implementation.
\vspace{0.1cm}

We use the \emph{doParallel} and \emph{foreach} packages in {\textsf{R} as developed and proposed by \cite{man:doParallel} and \cite{man:foreach}, see also
\cite{JSSv055i14}.
Since we have 100 financial firms we use the \emph{foreach} loops twice: the first loop is for the
100 financial firms with the second loop nested in the first loop to perform the moving window estimation. The speed of computation is increased considerably,
the script is available from Quantnet:
\raisebox{-1.5pt}{\includegraphics[scale=0.2]{Figures/QLogo}}\,{\href{https://github.com/QuantLet/FRM/tree/master/FRM_parallel_compute}{FRM\_parallel\_compute}}.
\vspace{0.1cm}

Without the use of parallel computing, i.e. using a processor with four cores for each moving window, it requires around two minutes to generate the FRM estimate
for one day. The Research Data Center (RDC) of Humboldt-Universit\"{a}t zu Berlin has provided access to their multi-core servers. Their servers have
respectively 24, 32, and 40 cores. By using these servers combined with parallel computing, the average computation time is reduced approximately $12$ seconds to
obtain a daily value for the FRM. The FRM webpage is updated weekly, which takes only 1 minute to generate the FRM series for five working days.
\vspace{0.1cm}

%%%%%%%%%%%%%%%%%%%%%%%%%%%%%%%%%%%%%%%%%%%%%%%%%%%%%%%%%%%%%%%%%%%%%%%%%%%%%%%%%%%%%%%%%%
\subsection{Visualization}
To implement the visualization of the FRM, we use the JavaScript framework
%The \emph{D3.js} library allows to create dynamic graphs.
D3.js (or just D3 for Data-Driven Documents), which is a JavaScript library for producing dynamic, interactive data visualizations in web browsers.
The \href{http://www.quantlet.de}{QuantNetXploRer} is a good example of D3 in power.
More information about the D3 architecture, its various designs and the D3-based QuantNetXploRer can be found in
\cite{Bostock2011_D3} and \cite{Q3D3LSA:BigDataAnalytics}.

\begin{figure}[ht]
	\centering
		\includegraphics[width=1.0\textwidth]{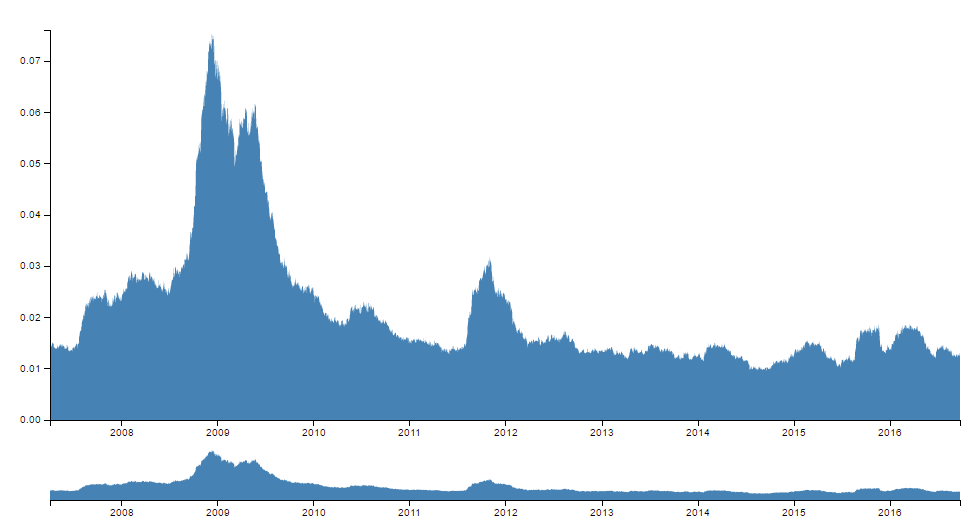}
		\caption{The graph of {\href{http://frm.wiwi.hu-berlin.de/}{Financial Risk Meter (FRM)}}.}
		\label{f1}
		\vspace{-0.5cm}
		\begin{flushright}
			\raisebox{-1.5pt}{\includegraphics[scale=0.2]{Figures/QLogo}}\,{\href{https://github.com/QuantLet/FRM/tree/master/FRM_parallel_compute}{FRM\_parallel\_compute}}
		\end{flushright}
\end{figure}

Figure \ref{f1} illustrates the D3-based FRM visualization, and more examples e.g. for Asia, Europe are available on \cite{Althof2019}.

%%%%%%%%%%%%%%%%%%%%%%%%%%%%%%%%%%%%%%%%%%%%%%%%%%%%%%%%%%%%%%%%%%%%%%%%%%%%%%%%%%%%%%%%%%
\subsubsection{Descriptive statistics}
Figure \ref{f1} shows the FRM series from April 5, 2007 through September 23, 2016. The FRM has no theoretical upper bound. In the time frame under consideration,
the maximum value is $0.075$, which occurred on December 15, 2008 and the mean value is $0.021$. We observe several peaks in the FRM series, which correspond to
crises and other events in these periods. Two peaks correspond to the financial crises in 2008 and 2010. The peak in the first quarter of 2009 is at the height of
the Great Recession: 800 thousand jobs were lost and the unemployment rate rose to $7.8\%$ in the US, which was the highest since June 1992. Another peak around
the fourth quarter of 2011 coincides with the decline in stock markets in August 2011, which was due to fears of contagion of the European sovereign debt crisis
to Spain and Italy.
\vspace{0.1cm}

Therefore one may state that the peaks of FRM series identify financial events and their impact on financial and systemic risk. The minimum of the FRM series in the time period under
consideration is observed in August 26, 2014, with a value of $0.009$. This was a relatively stable period.

%%%%%%%%%%%%%%%%%%%%%%%%%%%%%%%%%%%%%%%%%%%%%%%%%%%%%%%%%%%%%%%%%%%%%%%%%%%%%%%%%%%%%%%%%%
\subsubsection{Risk levels}
\label{seq:Risklevels}
For convenience and following the color scheme of US homeland security office we divide risk into five levels with different classifications and colors. The levels of risk are defined as different intervals of ratios for the FRM.
These ratios are computed based on the past values of the FRM. As shown in Figure \ref{f5}, we have five levels of risk with five color codes. The current risk level is determined by the
ratio based on all past FRM observations into which the current $\lambda$ falls. Table \ref{tquantile} presents the risk levels as well as the colors,
descriptions and ratios of the risk levels.
%\vspace{0.1cm}

\begin{figure}[ht]
	\centering
		\includegraphics[scale=0.8]{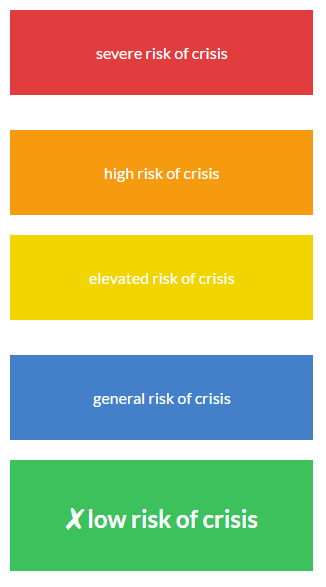}
		\caption{{\href{http://frm.wiwi.hu-berlin.de/}{Risk levels of FRM}}}
		\label{f5}
\end{figure}

\begin{table}[!ht]
\begin{center}
	\begin{tabular}{llc}
		\hline
		\hline
		Color & Risk level description & FRM ratio\\
    \hline
		Green   & Low risk of crisis in the financial market. & $<$20 \\
            & The incidence of a crisis is less likely than usual. & \\
		\hline
		Blue    & General risk of crisis in the financial market.  & 20-40 \\
            & There is no specific risk of a crisis. & \\
		\hline
		Yellow  & Elevated risk of crisis in the financial market.  & 40-60 \\
            & The incidence of a crisis is somewhat higher than usual. & \\
		\hline
		Orange  & High risk of crisis in the financial market.  & 60-80 \\
            & A crisis might occur very soon. & \\
		\hline
		Red     & Severe risk of a crisis in the financial market.  & $>$80 \\
            & A financial crisis is imminent or happening right now. & \\
		\hline
	\end{tabular}
	\caption{Risk levels, color codes and ratios for FRM}
	\label{tquantile}
\end{center}
\end{table}

As an example, on September 23, 2016 the value of FRM was $0.013$. Since the maximum of FRM series up to that date was $0.075$, the ratio of this risk
measure on September 23, 2016 was $17.3\%$. Since this is less than the 20\%-ratio, we classify the risk on that day as low risk of crisis in the financial
market with color green. On the website the current risk level is marked with a cross as shown in Figure \ref{f5} for this example.\\

%%%%%%%%%%%%%%%%%%%%%%%%%%%%%%%%%%%%%%%%%%%%%%%%%%%%%%%%%%%%%%%%%%%%%%%%%%%%%%%%%%%%%%%%%%
%%%%%%%%%%%%%%%%%%%%%%%%%%%%%%%%%%%%%%%%%%%%%%%%%%%%%%%%%%%%%%%%%%%%%%%%%%%%%%%%%%%%%%%%%%
\section{Causality of FRM and other systemic risk measures}
\cite{SFB649DP2016-047} analyze the factors affecting the value of $\lambda$ and summarize that $\lambda$ depends on three major factors: the variance of the error term,
the correlation structure of the covariates and the number of non-zero coefficients of the model. Since high volatility indicates high risk in finance and the number
of non-zero coefficients is related to the connectedness of the financial firms, they provide more theoretical evidence for the FRM as a risk measure. In their
application, they find a co-integration relationship between $\hat \lambda$ and other systemic risk measures. We extend their idea and use Granger causality
analysis to validate FRM. We select three measures: VIX \citep[see][]{Norman2009}, SRISK \citep[see][]{Brownlees:2016} as well as the
Google trends of the key word "financial crisis" \citep[see][]{GT2013}.
\vspace{0.1cm}

For the causality analysis we first need to introduce the Vector Autoregression (VAR) model briefly. \cite{Helmut2005} proposes the VAR(P) model as follows:\\
\begin{equation} \label{model_2}
 y_t= \alpha + A_1 y_{t-1} + A_2y_{t-2}+\cdots + A_Py_{t-P}+ {\mathbf{u}_t},
\end{equation}
where $y_t \stackrel{def}{=} (y_{1t}, \ldots, y_{Kt})^{\top}$, $A_i$ are fixed ($K \times K$) coefficient matrices,
$\mathbf{u}_t$ is a  $K$ dimensional process. The coefficients could be estimated by applying multivariate least squares estimation.
In order to perform the Granger causality test, the vector of endogenous variables $y_t$ is split into two subvectors $y_{1t}$ and $y_{2t}$ with dimensions
$(K_1 \times 1)$ and $(K_2 \times1)$ and $K = K_1 + K_2$. Then the VAR(P) model can be rewritten as follows:\\
\begin{eqnarray}
y_t &=&\left(
       \begin{array}{c}
         y_{1t} \\
         y_{2t}  \\
       \end{array}
     \right)
     =
 \left(
        \begin{array}{c}
          \alpha_1\\
          \alpha_2\\
        \end{array}
      \right)
      + \left(
         \begin{array}{cc}
           A_{11,1} & A_{12,1}  \\
           A_{21,1}  & A_{22,1}  \\
         \end{array}
       \right)
       \left(
         \begin{array}{c}
           y_{1,{t-1}} \\
           y_{2,{t-1}} \\
         \end{array}
       \right)
       + \cdots \nonumber \\&&
      \hspace{20ex}+
       \left(
         \begin{array}{cc}
           A_{11,P} & A_{12,P}  \\
           A_{21,P}  & A_{22,P}  \\
         \end{array}
       \right)
       \left(
         \begin{array}{c}
           y_{1,{t-P}} \\
           y_{2,{t-P}} \\
         \end{array}
       \right)
       +\left(
          \begin{array}{c}
            u_{1t} \\
            u_{2t} \\
          \end{array}
        \right)
\end{eqnarray}
The null hypothesis of the Granger causality test is that the subvector $y_{1t}$ does not Granger-cause $y_{2t}$, which is defined as $A_{21,i} = 0$ for
$i = 1, 2, \ldots , P$. The alternative hypothesis states that the subvector $y_{1t}$ Granger-causes $y_{2t}$ and is defined as: $\exists \, A_{21,i} \neq 0$
for $i = 1, 2, \ldots , P$. The test statistic follows an $F$ distributions with  $PK_1K_2$ and  $KJ - n^*$ degrees of freedom, where $J$ is the sample size and
$n^*$ equals the total number of parameters in the above VAR(P) model.
\vspace{0.1cm}

\subsection{FRM versus VIX}\label{section_FRM_VIX}
The VIX series is often addressed as a ``fear index'' and can be interpreted as a measure for systemic risk \citep{Norman2009}. For reasons of comparability, we
standardize these two series by setting the lowest value in the sample to zero and the highest to one. Figure \ref{Figure_FRM_VIX} plots the standardized FRM series
(thick black line) and the VIX series (thin red line). The plot shows that both indicators move in the same direction, with the VIX series a little more volatile.
We also get some evidence of some financial events by observing the corresponding volatility levels of the FRM and VIX. For example, in the end of 2008 there is a
sharp upward trend of FRM, whereas the upward trends dominates VIX as well, which corresponds to the bankruptcy of Lehman Brothers on September 15, 2008.
Both FRM and VIX have higher values between 2008 and 2010, which corresponds to the time period of the financial crises. After 2013 the values of FRM are relative
stable at a low level, while there is similar pattern of VIX, which shows signs of the slow recovery of the global economy from the recession.

\begin{figure}[!ht]
	\begin{center}
		\includegraphics[width=0.9\textwidth]{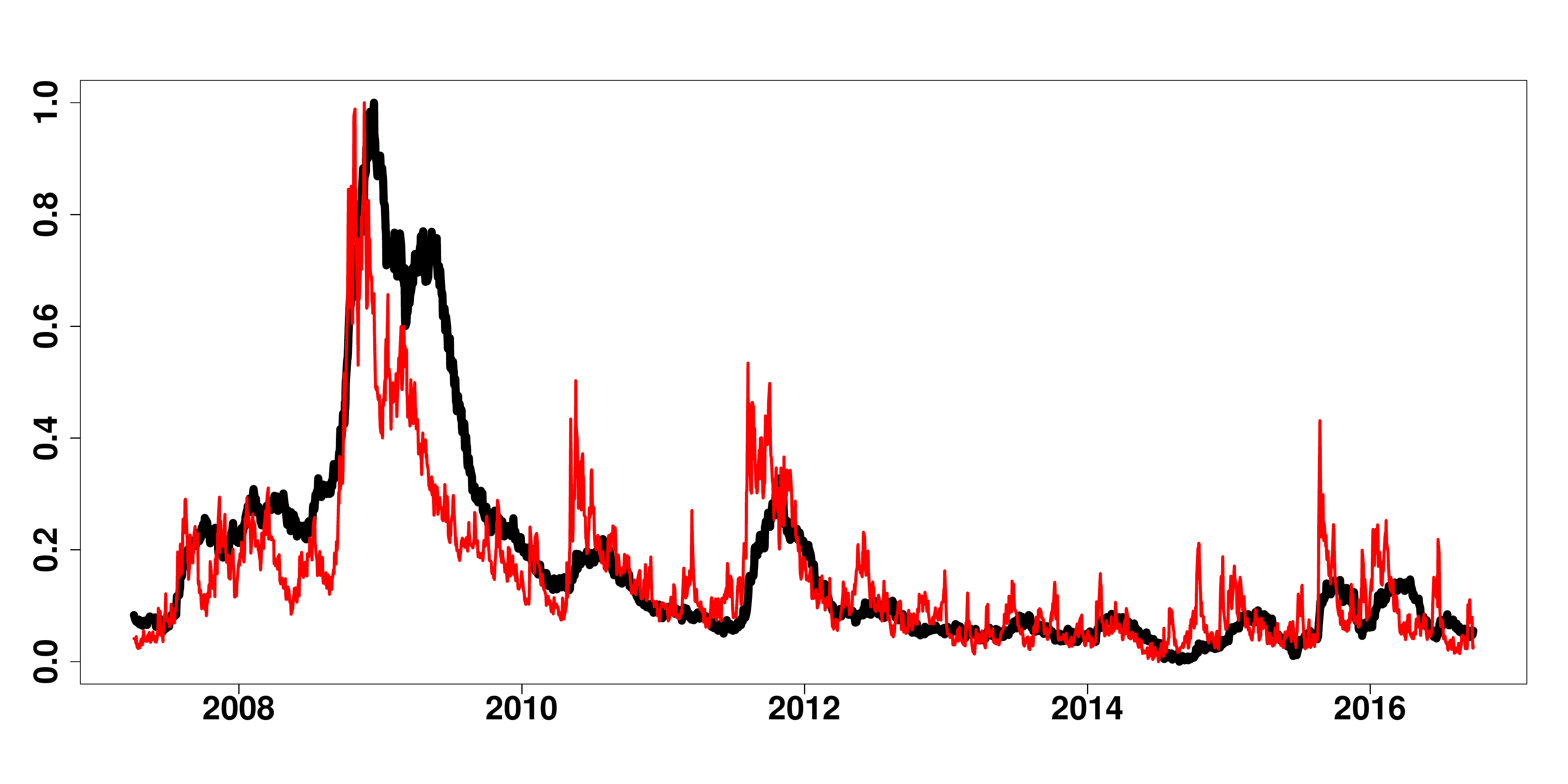}
		\caption{Scaled FRM (thick black line) and VIX (thin red line)}
		\label{Figure_FRM_VIX}
	\end{center}
	\vspace{-0.5cm}
	\begin{flushright}
		\raisebox{-1.5pt}{\includegraphics[scale=0.2]{Figures/QLogo}}\,{\href{https://github.com/QuantLet/FRM/tree/master/FRM_VIX}{FRM\_VIX}}
	\end{flushright}
\end{figure}

Before we perform the Granger causality test, we test for stationarity of both time series with the Augmented Dickey-Fuller (ADF) test. The results of the test are shown in Table \ref{tb_1}. For the FRM series, the $p$ value is larger
than $0.05$, so we cannot reject the null hypothesis, i.e. the FRM series may have a unit root and may be non-stationary. We reject the null hypothesis for the VIX series
with a $p$ value smaller than $0.05$ and conclude that the VIX series is stationary.  We do not need to consider the co-integration problem, since only if both
series are non-stationary, we should take into account the co-integration. There is a trade-off between using the original data and the transformed (differenced)
data to find the causality relationship. \cite{Sims1980} prefers to use the original data. He argues that VAR with non-stationary variables may provide important
insights, if one is interested in the nature of relationships between variables. \cite{Brooks2014} also states that differencing will destroy information on any
long-run relationships between the series. However, other people argue that the original non-stationary data might lead to untrusted estimation, see \cite{Yule1926}
and \cite{Granger1974}. In our case, we consider both the original data and transformed data.
\vspace{0.1cm}

\renewcommand\arraystretch{1.5}
\begin{table}[!ht]
	\begin{center}
		\begin{tabular}{cc}
			\hline\hline
			Series & $p$ values \\
			\hline
			FRM&$0.28$\\
			VIX &0.01\\
			DFRM&$0.01$\\
			\hline\hline
		\end{tabular}
		\caption{$p$ values of ADF test for stationarity}
		\label{tb_1}
	\end{center}
	\vspace{-0.5cm}
	\begin{flushright}
		\raisebox{-1.5pt}{\includegraphics[scale=0.2]{Figures/QLogo}}\,{\href{https://github.com/QuantLet/FRM/tree/master/FRM_VIX}{FRM\_VIX}}
	\end{flushright}
\end{table}

\renewcommand\arraystretch{1.5}
\begin{table}[!ht]
	\begin{center}
		\begin{tabular}{ccccc}
			\hline\hline
			Model & AIC& HQ & SC & FPE\\
			\hline
			FRM and VIX &20& $3$& $3$& 20\\
			DFRM and VIX  &19& $8$& $5$& 19\\
			\hline\hline
		\end{tabular}
		\caption{Suggested order for VAR process by different criteria}
		\label{tb_4}
	\end{center}
\end{table}

\renewcommand\arraystretch{1.5}
\begin{table}[!ht]
	\resizebox{\columnwidth}{!}{
		\begin{tabular}{crrrrr}
			\hline\hline
			Model & Order VAR & PT (asymptotic)& PT (adjusted) & BG & ES\\
			\hline
			\multirow{2}{*}{FRM and VIX}& 3 &$< 2.2 \times 10^{-16}$&$< 2.2 \times 10^{-16}$&  $1.1\times10^{-07}$&  $1.0\times10^{-07}$\\
			\cline{2-6}
			& 11 &$2.5 \times 10^{-07}$&$2.0\times 10^{-07}$&  $1.6\times10^{-01}$&  $1.7\times10^{-01}$\\
			\cline{2-6}
			& 20 &$< 2.2 \times 10^{-16}$&$< 2.2 \times 10^{-16}$&  $3.1\times10^{-08}$&  $4.1\times10^{-08}$\\
			\cline{2-6}
			\hline
			\multirow{4}{*}{DFRM and VIX}&  5 &$2.2 \times 10^{-16}$&$2.2 \times 10^{-16}$&  $3.2\times10^{-08}$&  $ 3.1\times10^{-08}$\\
			\cline{2-6}
			& 8 &$6.7 \times 10^{-12}$ &$4.9\times 10^{-12}$ &$1.4\times 10^{-06}$ &$1.5\times 10^{-06}$\\
			\cline{2-6}
			& 11 &$ 2.3 \times 10^{-09}$ &$1.8\times 10^{-09}$ &$1.5\times 10^{-03}$ &$1.7\times 10^{-03}$\\
			\cline{2-6}
			& 19 &$1.7 \times 10^{-03}$ &$1.6 \times 10^{-03}$ &$5.5\times 10^{-08}$ &$7.2\times 10^{-08}$\\
			\hline\hline
		\end{tabular}}
	\caption{$p$ values of model selection tests}
	\label{tb_10}
\end{table}
%\vspace{0.5cm}

\begin{table}[!ht]
	\begin{center}
		\begin{tabular}{lll}
			\hline\hline
			Cause & Effect & $p$ values\\
			\hline
			FRM &VIX& $4.0 \times 10^{-08}$\\
			\hline
			VIX &FRM & $6.1 \times 10^{-11}$\\
			\hline
			DFRM &VIX & $6.6 \times 10^{-11}$\\
			\hline
			VIX &DFRM & $8.7 \times 10^{-13}$ \\
			\hline\hline
		\end{tabular}
		\caption{$p$ values of Granger causality test}
		\label{tb_6}
	\end{center}
	\vspace{-0.5cm}
	\begin{flushright}
		\raisebox{-1.5pt}{\includegraphics[scale=0.2]{Figures/QLogo}}\,{\href{https://github.com/QuantLet/FRM/tree/master/FRM_VIX}{FRM\_VIX}}
	\end{flushright}
\end{table}

\begin{figure}[!ht]
	\centering
	\includegraphics[width=0.62\textwidth]{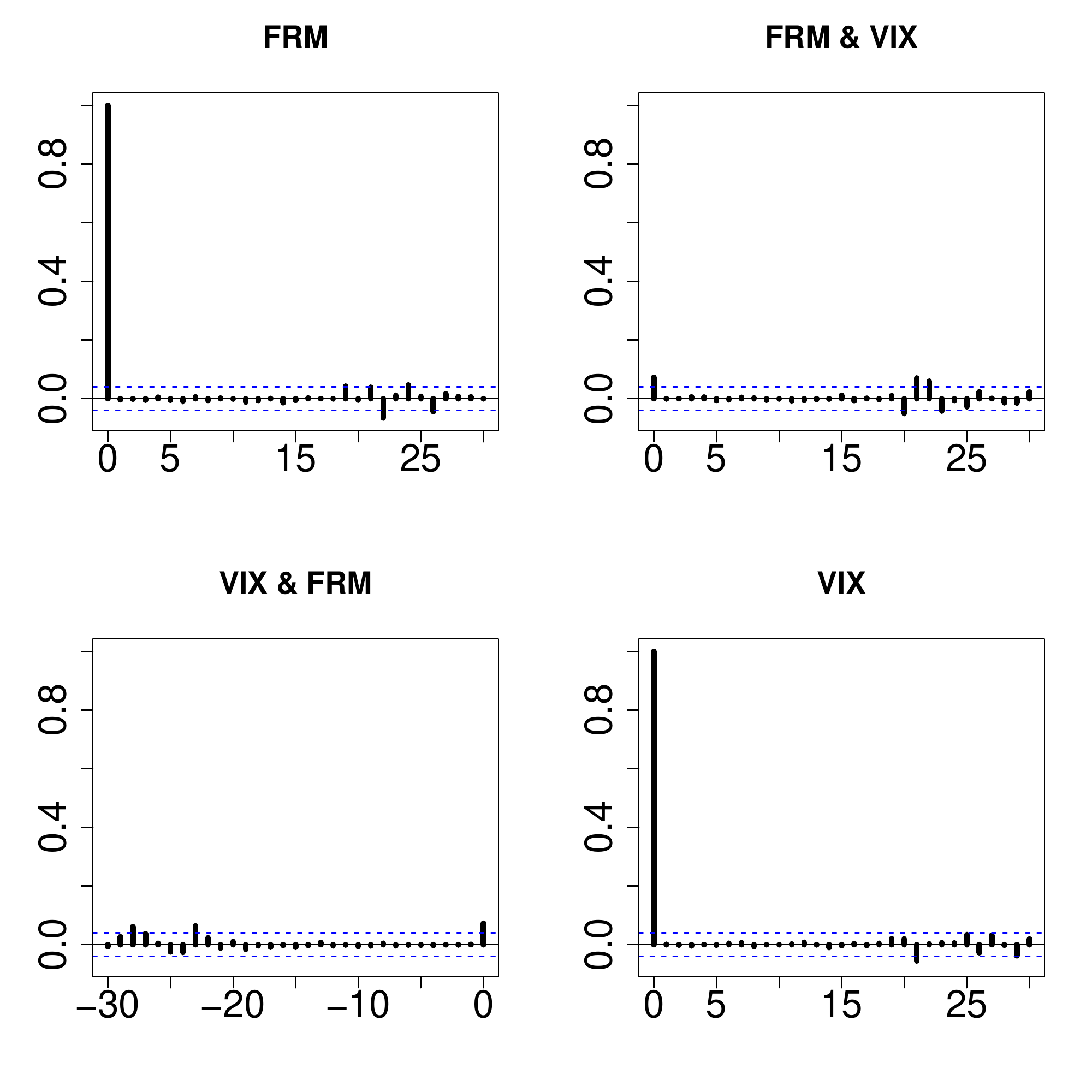}
	\caption{Autoregression functions of FRM and VIX}
	\label{f3}
	\vspace{-0.5cm}
	\begin{flushright}
		\raisebox{-1.5pt}{\includegraphics[scale=0.2]{Figures/QLogo}}\,{\href{https://github.com/QuantLet/FRM/tree/master/FRM_VIX}{FRM\_VIX}}
	\end{flushright}
\end{figure}

\begin{figure}[!ht]
	\centering
	\includegraphics[width=0.62\textwidth]{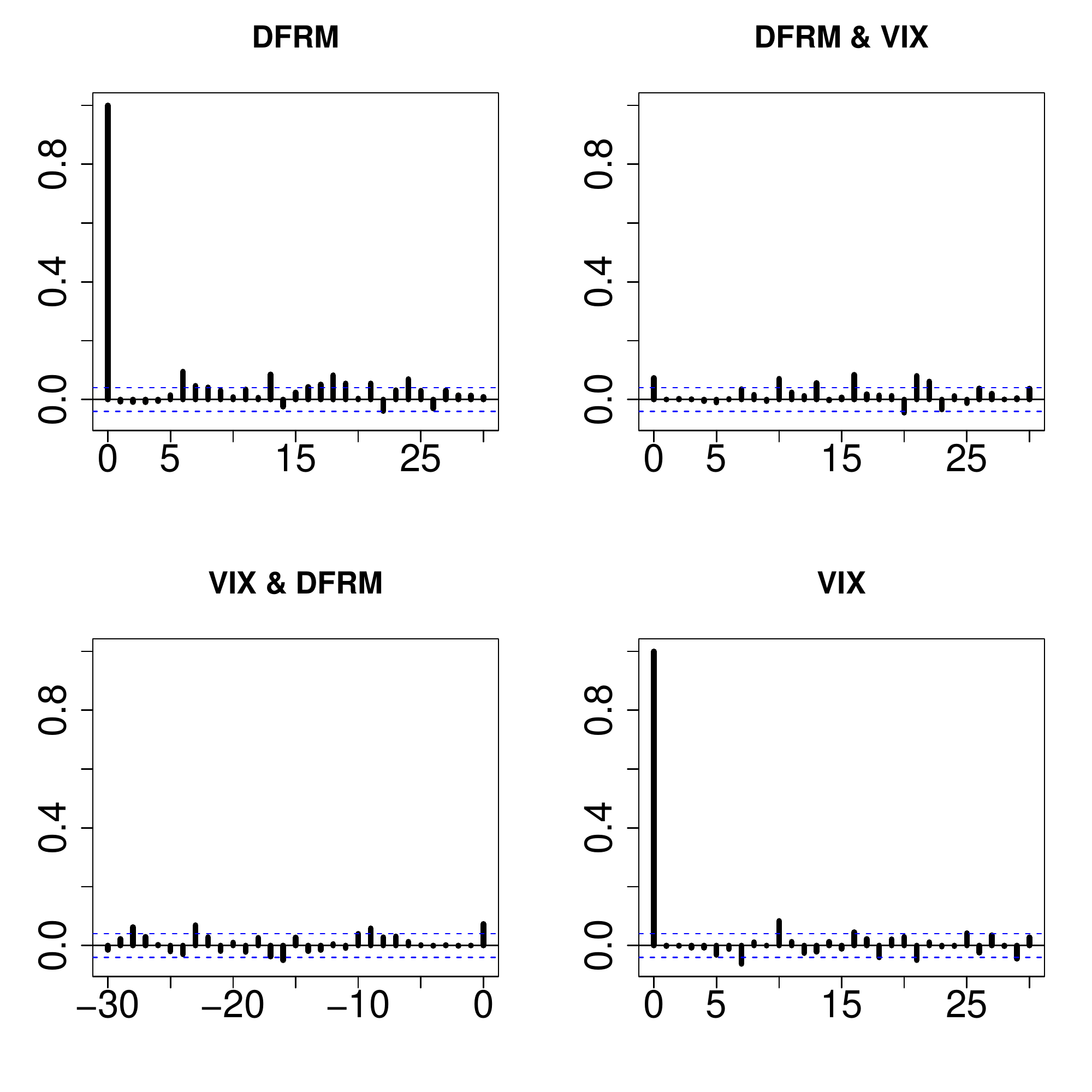}
	\caption{Autoregression functions of DFRM and VIX}
	\label{f4}
	\vspace{-0.5cm}
	\begin{flushright}
		\raisebox{-1.5pt}{\includegraphics[scale=0.2]{Figures/QLogo}}\,{\href{https://github.com/QuantLet/FRM/tree/master/FRM_VIX}{FRM\_VIX}}
	\end{flushright}
\end{figure}

Firstly, we consider the original data. We choose the VAR order according to four criteria: the Akaike information criterion (AIC), the Hannan-Quinn information
criterion\\\\(HQ), the Schwarz criterion (SC) and the Prediction Error Criterion (FPE), see Table \ref{tb_4}. While HQ and SC suggest an order 3 VAR process, AIC and
FPE suggest an order 20 process. We fit both VAR models with order 3 and order 20.  Next, we check the autocorrelation of the residuals to decide the optimal order.
Four tests are carried out: the asymptotic Portmanteau Test, the adjusted Portmanteau Test, the Breusch-Godfrey LM test and the Edgerton-Shukur F test. The null
hypothesis of these tests is that there is no first order autocorrelation among residuals. Choosing order 3 and 20 leads to the rejection of all these tests
(cf. Table \ref{tb_10}). Subsequently we try the other orders and find that with order 11 both the Breusch-Godfrey LM test and the Edgerton-Shukur F tests are passed.
Therefore, we select order 11. The autocorrelation function of the residuals is plotted in Figure \ref{f3}. Table \ref{tb_6} shows the results of the Granger
causality test. All $p$ values are smaller than $0.05$ which indicates that the null hypothesis is rejected. Therefore, FRM Granger causes VIX, and also VIX Granger
causes FRM.
\vspace{0.1cm}

Next, we consider the transformed series. Since FRM is non-stationary, we take the first difference. The transformed series is called as DFRM. In Table \ref{tb_1}
we see that DFRM is stationary. Then the same procedure as before is performed. While HQ suggests an order 8 process, SC suggest an order 5, and AIC and FPE both
suggest an order 19 (cf. Table \ref{tb_4}). After checking the four tests for autocorrelation of the residuals, we conclude that the optimal order is 19. Although
it does not pass the autocorrelation test, the $p$ value is close to the critical value $0.05$, and the autocorrelation function confirms this result
(cf. Table \ref{tb_10} and Figure \ref{f4}). The result of the Granger causality test is summarized in Table \ref{tb_6}. We find that all $p$ values are
significantly smaller than $0.05$, which indicates that the null hypothesis is rejected. Therefore we conclude that DFRM Granger causes VIX, and also VIX Granger
causes DFRM.

%%%%%%%%%%%%%%%%%%%%%%%%%%%%%%%%%%%%%%%%%%%%%%%%%%%%%%%%%%%%%%%%%%%%%%%%%%%%%%%%%%%%%%%%%%
\subsection{FRM versus SRISK}

SRISK is a macro-finance measure of systemic risk \citep{Acharya, Brownlees:2016}. Our data on SRISK for the US are obtained from V-Lab
\footnote{See the Systemic Risk Analysis Welcome Page: https://vlab.stern.nyu.edu/welcome/risk/}. We also standardize SRISK, so that both series are comparable on
the same scale. Figure \ref{Figure_FRM_SRISK} plots the standardized FRM series (thick black line) and the SRISK series (thin blue line). We see that there is a peak
in the first quarter of 2008 for SRISK, but afterwards FRM and SRISK have similar patterns. Especially during the beginning of 2010 and the beginning of 2012,
the two series have a similar shape.

\begin{figure}[!ht]
	\begin{center}
		\includegraphics[width=0.9\textwidth]{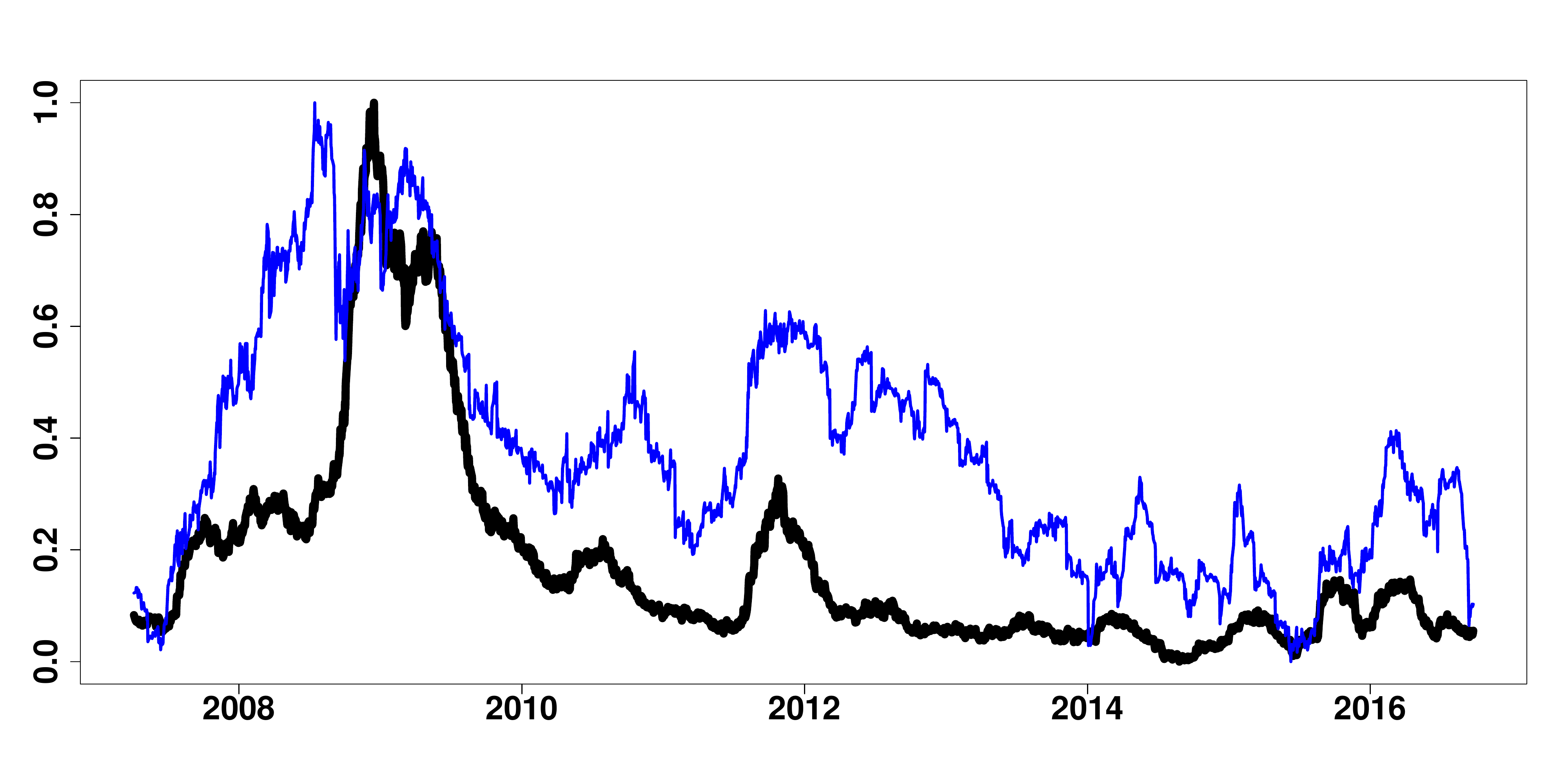}
		\caption{Scaled FRM (thick black line) and SRISK (thin blue line)}
		\label{Figure_FRM_SRISK}
	\end{center}
	\vspace{-0.5cm}
	\begin{flushright}
		\raisebox{-1.5pt}{\includegraphics[scale=0.2]{Figures/QLogo}}\,{\href{https://github.com/QuantLet/FRM/tree/master/FRM_SRISK}{FRM\_SRISK}}
	\end{flushright}
\end{figure}

\renewcommand\arraystretch{1.5}
\begin{table}[!ht]
	\begin{center}
		\begin{tabular}{cc}
			\hline\hline
			Variables & p-values\\
			\hline
			FRM & $0.48$\\
			SRISK & 0.10\\
			\hline\hline
		\end{tabular}
	\end{center}
	\caption{$p$ values of ADF test for stationarity for FRM and SRISK}
	\label{tb_7}
\end{table}

\begin{table}[!ht]
	\begin{tabular}{cccc}
		\hline\hline
		Explanatory (Cause) & Response (Effect) & Value of test-statistic & Critical value at 5\%\\
		\hline
		FRM & SRISK & -3.1& -1.95\\
		\hline
		SRISK & FRM & -2.7& -1.95\\
		\hline\hline
	\end{tabular}
	\caption{Results of Engle Granger 2-step co-integration test}
	\label{tb_11}
	\vspace{-0.5cm}
	\begin{flushright}
		\raisebox{-1.5pt}{\includegraphics[scale=0.2]{Figures/QLogo}}\,{\href{https://github.com/QuantLet/FRM/tree/master/FRM_SRISK}{FRM\_SRISK}}
	\end{flushright}
\end{table}

We perform the same procedure as in section \ref{section_FRM_VIX}. The results of the ADF test for the SRISK series in Table \ref{tb_7} show that the series is
non-stationary. Since the FRM series is neither stationary, we consider the co-integration of them. From \cite{GRANGER1988199} we know that if both series are
co-integrated, then there must be Granger causality between them in at least one way. We perform the Engle Granger 2-step test for co-integration, which is suitable
for bivariate time series. In the first step, the linear regression of FRM on SRISK is carried out, i.e. FRM is the explanatory variable and SRISK the response
variable. In the second step, we test the residuals of the aforementioned linear regression. If these residuals are stationary, then there is co-integration of FRM
and SRISK. The null hypothesis of this test is that the residuals are non-stationary. The result of this test are summarized in Table \ref{tb_11}. We conclude that
FRM and SRISK are co-integrated, in other words, FRM Granger causes SRISK. If we regress SRISK on FRM, i.e. SRISK is the explanatory variable and FRM the response
variable, we also conclude that SRISK and FRM are co-integrated, which indicates that SRISK Granger causes FRM. We thus conclude that there is mutual causality
between FRM and SRISK.

\clearpage

%%%%%%%%%%%%%%%%%%%%%%%%%%%%%%%%%%%%%%%%%%%%%%%%%%%%%%%%%%%%%%%%%%%%%%%%%%%%%%%%%%%%%%%%%%
\subsection{FRM versus Google Trends}
Finally, we analyze the relationship between FRM and Google Trends (GT) for the keyword "financial crisis".
Google Trends provides data on the search volume of particular words and phrases relative to the total search volume. This can be disaggregated by countries.
If a keyword is more frequently searched for, this might indicate a particular interest. \cite{GT2013} analyzed the data related to finance from Google Trends,
and find that Google Trends data did not only reflect the current state of the stock markets, but may have also been able to forecast certain future trends.
We use Google Trends for the keyword "financial crisis", assuming that more people will search for this term if they feel the risk for a financial crisis is high.
The Google Trends data are weekly data. To allow for comparison with the FRM we apply cubic interpolation to estimate daily data from the weekly Google Trends series.
This series is compared with the daily FRM series. Figure \ref{Figure_FRM_gt} plots both the daily FRM series as well as the cubic interpolated Google Trends daily
series. Both series are standardized to the interval zero-one for comparison. We observe some co-movement between both series, but no continuous superiority of the Google Trends above FRM.

\begin{figure}[ht]
	\centering
	\includegraphics[width=0.9\textwidth]{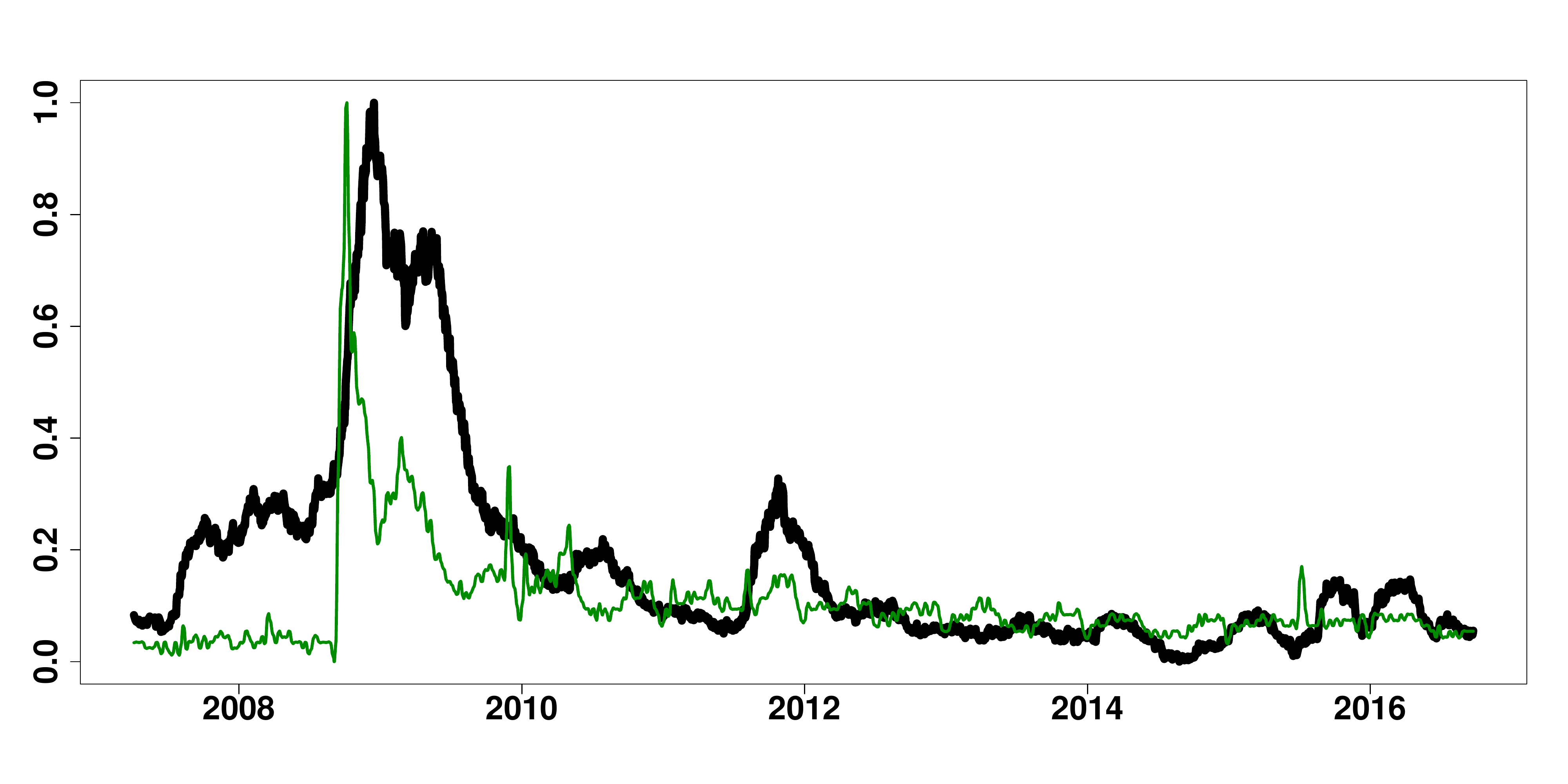}
	\caption{Scaled FRM (thick black line) and Google Trends (thin green line)}
	\label{Figure_FRM_gt}
	\vspace{-0.5cm}
	\begin{flushright}
		\raisebox{-1.5pt}{\includegraphics[scale=0.2]{Figures/QLogo}}\,{\href{https://github.com/QuantLet/FRM/tree/master/FRM_GT}{FRM\_GT}}
	\end{flushright}
\end{figure}

The ADF test shows that the GT series is stationary (cf. Table \ref{tb_2}). We  perform two tests for the relationship between the two series. Firstly, we consider
the original data of FRM, then we consider the transformed data. We perform four criteria to find the optimal order of VAR model. As the results in Table \ref{tb_8}
show, all the criteria suggest an order 20 VAR process. Therefore, we apply an order 20 VAR model. Next, the autocorrelation of the residuals is tested. Although
none of the tests can be passed (cf. Table \ref{tb_5}), we have no better choice for the order than 20. The autocorrelation function of residuals are plotted in
Figure \ref{f8}. Table \ref{tb_9} shows the results of the Granger causality test. All $p$ values are significantly smaller than $0.05$, which indicates that the
null hypothesis is rejected. Therefore, FRM Granger causes GT, and GT Granger causes FRM.
\vspace{0.1cm}

For the first differenced FRM, i.e. DFRM, the same procedure is used. In Table \ref{tb_8} all the criteria suggest an order 20 VAR process. The result of the
autocorrelation tests are presented in Table \ref{tb_8}. Although none of the tests is passed, we still use order 20. The autocorrelation function of the residuals
is shown in Figure \ref{f9}. Table \ref{tb_9} shows the results of the Granger causality test. All $p$ values are significantly smaller than $0.05$, which indicates
that the null hypothesis is rejected. Therefore, DFRM Granger causes GT, and GT Granger causes DFRM.

\begin{table}[ht]
	\begin{center}
		\begin{tabular}{cc}
			\hline\hline
			Variables &p-values \\
			\hline
			FRM&$0.48$\\
			GT &0.01\\
			DFRM&$0.01$\\
			\hline\hline
		\end{tabular}
		\caption{$p$ values of ADF test for stationarity for FRM and GT}
		\label{tb_2}
	\end{center}
\end{table}

\begin{table}[ht]
	\begin{center}
		\begin{tabular}{ccccc}
			\hline\hline
			Model & AIC& HQ & SC & FPE\\
			\hline
			FRM and GT &20& $20$& $20$& 20\\
			DFRM and GT &20& $20$& $20$& 20\\
			\hline\hline
		\end{tabular}
		\caption{Suggested order for VAR process by different criteria}
		\label{tb_8}
	\end{center}
\end{table}

\vspace{0.5cm}
\renewcommand\arraystretch{1.5}
\begin{table}[!ht]
	\resizebox{\columnwidth}{!}{
	\begin{tabular}{cccccc}
		\hline\hline
		Model & Order & PT (asymptotic)& PT (adjusted) & BG & ES\\
		\hline
		FRM and GT&  20&$< 2.2 \times 10^{-16}$& $< 2.2 \times 10^{-16}$ &$< 2.2 \times 10^{-16}$&$< 2.2 \times 10^{-16}$\\
		\hline
		DFRM and GT& 20&$< 2.2 \times 10^{-16}$& $< 2.2 \times 10^{-16}$ &$< 2.2 \times 10^{-16}$&$< 2.2 \times 10^{-16}$\\
		\hline\hline
	\end{tabular}}
	\caption{$p$ values of model selection tests}
	\label{tb_5}
\end{table}

\vspace{0.5cm}
\begin{table}[!ht]
	\begin{center}
		\begin{tabular}{lll}
			\hline\hline
			Cause & Effect & p-values\\
			\hline
			FRM &GT& $ 1.1 \times 10^{-10}$ \\
			\hline
			GT &FRM & $2.1\times 10^{-12}$ \\
			\hline
			DFRM &GT & $6.8 \times 10^{-11}$ \\
			\hline
			GT &DFRM & $4.1 \times 10^{-10}$\\
			\hline\hline
		\end{tabular}
		\caption{$p$ values of Granger causality test}
		\label{tb_9}
	\end{center}
	\vspace{-0.5cm}
	\begin{flushright}
		\raisebox{-1.5pt}{\includegraphics[scale=0.2]{Figures/QLogo}}\,{\href{https://github.com/QuantLet/FRM/tree/master/FRM_GT}{FRM\_GT}}
	\end{flushright}
\end{table}

\begin{figure}[!ht]
	\centering
	\includegraphics[width=0.64\textwidth]{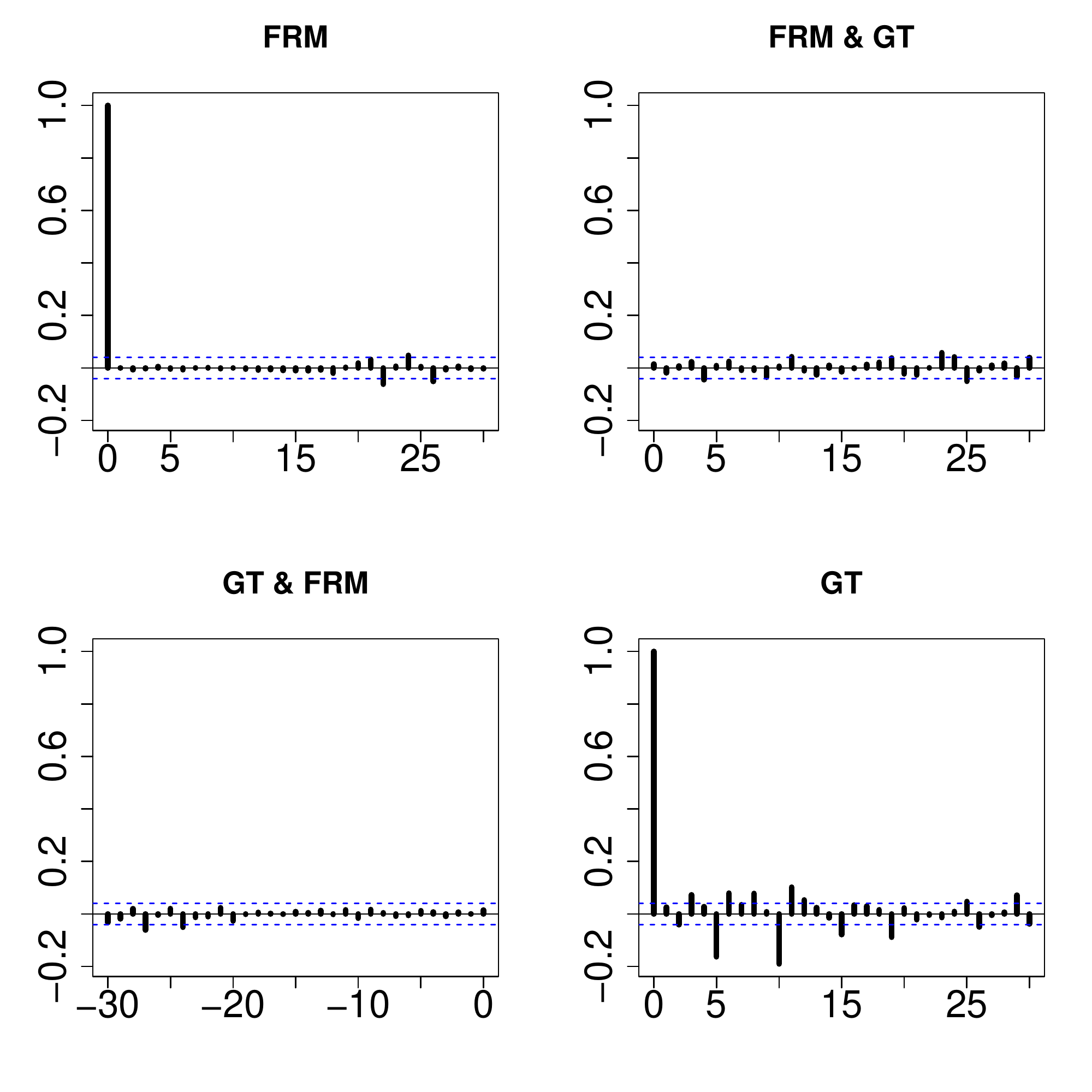}
	\caption{Autoregression functions of FRM and GT}
	\label{f8}
	\vspace{-0.5cm}
	\begin{flushright}
		\raisebox{-1.5pt}{\includegraphics[scale=0.2]{Figures/QLogo}}\,{\href{https://github.com/QuantLet/FRM/tree/master/FRM_GT}{FRM\_GT}}
	\end{flushright}
\end{figure}

\begin{figure}[!ht]
	\centering
	\includegraphics[width=0.64\textwidth]{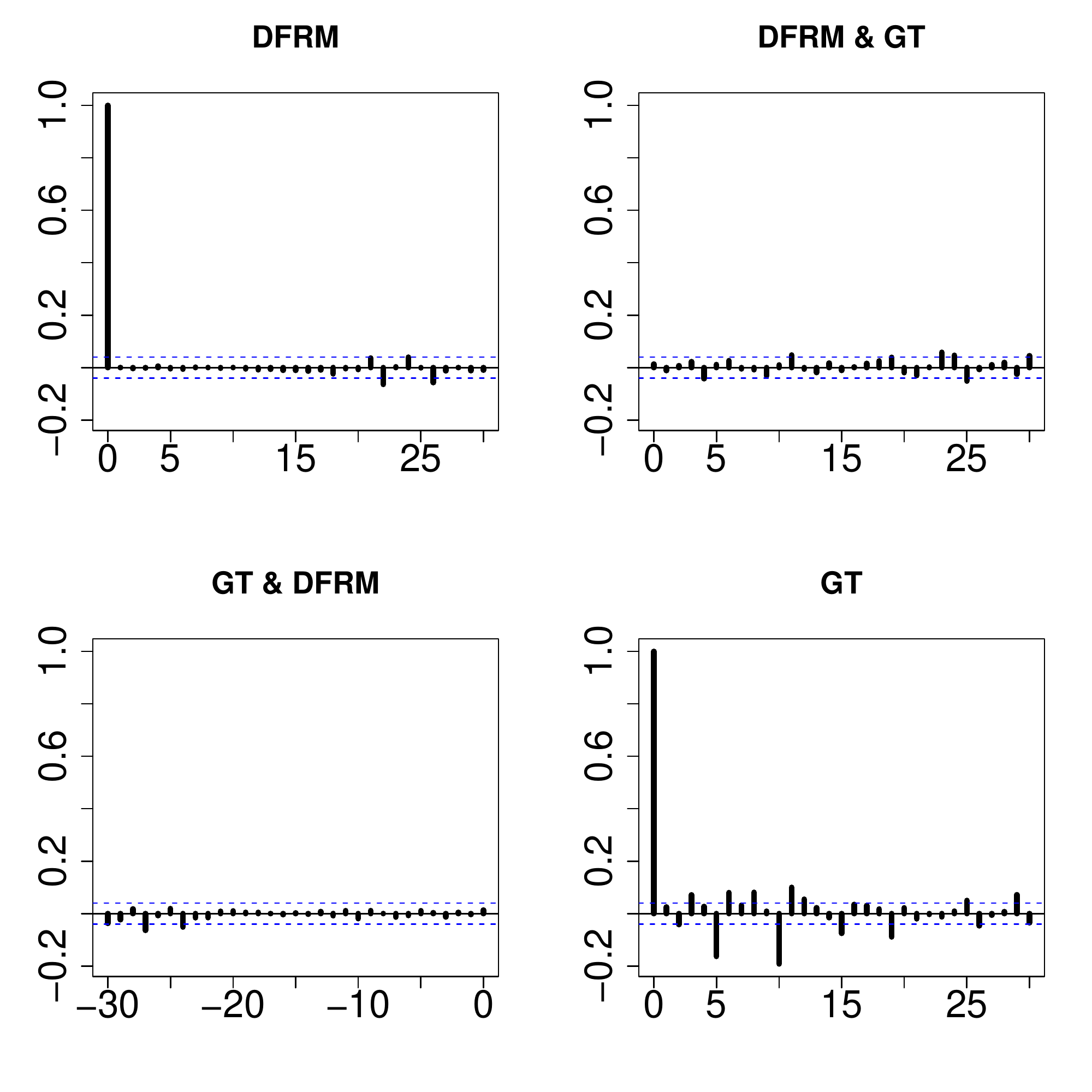}
	\caption{Autoregression functions of DFRM and GT}
	\label{f9}
	\vspace{-0.5cm}
	\begin{flushright}
		\raisebox{-1.5pt}{\includegraphics[scale=0.2]{Figures/QLogo}}\,{\href{https://github.com/QuantLet/FRM/tree/master/FRM_GT}{FRM\_GT}}
	\end{flushright}
\end{figure}

\clearpage

%%%%%%%%%%%%%%%%%%%%%%%%%%%%%%%%%%%%%%%%%%%%%%%%%%%%%%%%%%%%%%%%%%%%%%%%%%%%%%%%%%%%%%%%%%
%%%%%%%%%%%%%%%%%%%%%%%%%%%%%%%%%%%%%%%%%%%%%%%%%%%%%%%%%%%%%%%%%%%%%%%%%%%%%%%%%%%%%%%%%%
\section{Conclusion}
In this paper we propose and develop an AI based measure for systemic risk in financial markets: the Financial Risk Meter (FRM). The FRM is a measure for systemic risk based
on the penalty term $\lambda$ of the linear quantile lasso regression, which is defined as the average of the $\lambda$ series over the 100 largest US publicly
traded financial institutions. The implementation is carried out by using parallel computing. The risk levels are classified by five levels. The empirical result
shows that our Financial Risk Meter can be a good indicator for trends in systemic risk. Compared with other systemic risk measures, such as VIX, SRISK,
Google Trends with the keyword ``financial crisis'', we find that the FRM and VIX, FRM and SRISK, FRM and GT mutually granger cause one another, which means that
our FRM is a good measure of systemic risk for the US financial market. All the codes of FRM are published on \url{www.quantlet.de} with keyword
\raisebox{-1.5pt}{\includegraphics[scale=0.2]{Figures/QLogo}}\,{\href{https://github.com/QuantLet/FRM}{FRM}}.
The \textsf{R} package \textbf{RiskAnalytics} \citep{RiskAnalytics} is another tool with the purpose of integrating
and facilitating the research, calculation and analysis methods around the FRM project \citep{RiskAnalyticsDP}.
The up-to-date FRM can be found on \href{http://frm.wiwi.hu-berlin.de/}{hu.berlin/frm}.

\renewcommand\arraystretch{1.0}
\begin{landscape}
\footnotesize
\section{Appendix: Financial Institutions}
\begin{center}
\begin{longtable}[H]{ l l | l l }
  %\centering
    %\begin{tabular}
    \hline
     \hline
    %& \textbf{Depositories (25)} & &\textbf{Insurances (25)}  \\
&&&\\
&&&\\
      %  \hline
WFC&Wells Fargo \& Company&AON&Aon plc\\
JPM&J P Morgan Chase \& Co& ALL&Allstate Corporation\\
BAC&Bank of America Corporation&BEN&Franklin Resources, Inc.\\
C&Citigroup Inc.&STI&SunTrust Banks, Inc.\\
AIG&American International Group, Inc.&MCO&Moody's Corporation\\
GS&Goldman Sachs Group, Inc.&PGR&Progressive Corporation\\
USB&U.S. Bancorp&AMP&Ameriprise Financial Services, Inc.\\
AXP&American Express Company&AMTD&TD Ameritrade Holding Corporation\\
MS&Morgan Stanley&HIG&Hartford Financial Services Group, Inc.\\
BLK&BlackRock, Inc.&TROW&T. Rowe Price Group, Inc.\\
MET&MetLife, Inc.&NTRS&Northern Trust Corporation\\
%COF&Capital One Financial Corporation&CB&Chubb Corporation (The)\\
PNC&PNC Financial Services Group, Inc. (The)&MTB&M\&T Bank Corporation\\%MMC&Marsh \& McLennan Companies, Inc.\\
BK&Bank Of New York Mellon Corporation (The)&FITB&Fifth Third Bancorp\\
SCHW&The Charles Schwab Corporation&IVZ&Invesco Plc \\
COF&Capital One Financial Corporation&L&Loews Corporation\\
PRU&Prudential Financial, Inc.& EFX& Equifax, Inc.\\
TRV&The Travelers Companies, Inc.&PFG&Principal Financial Group Inc\\
CME&CME Group Inc.&RF&Regions Financial Corporation\\
CB&Chubb Corporation&MKL&Markel Corporation\\
MMC&Marsh \& McLennan Companies, Inc.&FNF&Fidelity National Financial, Inc.\\
BBT&BB\&T Corporation&LNC&Lincoln National Corporation\\
ICE&Intercontinental Exchange Inc.&CBG&CBRE Group, Inc.\\
STT&State Street Corporation&KEY&KeyCorp\\
AFL&Aflac Incorporated&NDAQ&The NASDAQ OMX Group, Inc.\\
CINF&Cincinnati Financial Corporation&CACC&Credit Acceptance Corporation\\
%&&&\\
&&&\\
\hline
&&&\\
%&&&\\
CNA&CNA Financial Corporation&BRO&Brown \& Brown, Inc.\\
HBAN&Huntington Bancshares Incorporated&ERIE&Erie Indemnity Company\\
SEIC&SEI Investments Company&OZRK&Bank of the Ozarks\\
ETFC&E*TRADE Financial Corporation&WTM&White Mountains Insurance Group, Ltd.\\
AMG&Affiliated Managers Group, Inc.&SNV&Synovus Financial Corp.\\
RJF&Raymond James Financial, Inc.& ISBC&Investors Bancorp, Inc.\\
UNM&Unum Group&MKTX&MarketAxess Holdings, Inc.\\
NYCB&New York Community Bancorp, Inc.&LM&Legg Mason, Inc. \\
Y&Alleghany Corporation&CBSH&Commerce Bancshares, Inc.\\
SBNY&Signature Bank&BOKF&BOK Financial Corporation\\
CMA&Comerica Incorporated&EEFT&Euronet Worldwide, Inc.\\
AJG&Arthur J. Gallagher \& Co.&DNB&Dun \& Bradstreet Corporation\\
TMK&Torchmark Corporation&WAL&Western Alliance Bancorporation\\
WRB&W.R. Berkley Corporation&EV&Eaton Vance Corporation\\
AFG&American Financial Group, Inc.&CFR&Cullen/Frost Bankers, Inc.\\
SIVB&SVB Financial Group&MORN&Morningstar, Inc.\\
EWBC&East West Bancorp, Inc.& THG&The Hanover Insurance Group, Inc.\\
ROL&Rollins, Inc.&UMPQ&Umpqua Holdings Corporation\\
ZION&Zions Bancorporation&CNO&CNO Financial Group, Inc.\\
AIZ& Assurant, Inc.& FHN&First Horizon National Corporation\\
PACW&PacWest Bancorp&WBS&Webster Financial Corporation\\
AFSI&AmTrust Financial Services, Inc.& PB & Prosperity Bancshares, Inc.\\
ORI&Old Republic International Corporation&PVTB&PrivateBancorp, Inc.\\
PBCT&People's United Financial, Inc.&SEB&Seaboard Corporation\\
FCNCA&First Citizens BancShares, Inc.&MTG&MGIC Investment Corporation\\
&&&\\
\hline
\hline
  \caption{The list of 100 financial companies used to estimate FRM in our sample.}
  \label{table:List_of_Firms}
\end{longtable}
\end{center}
\end{landscape}

\newpage

\addcontentsline{toc}{chapter}{References}
%\addcontentsline{toc}{section}{References}

\bibliography{bibliographyFRM}

\end{document}